\documentclass[useAMS,usenatbib,usegraphicx]{mn2e}

\usepackage{morefloats}
\usepackage{soul}
\usepackage{color}
\usepackage{float}

\title[Near and far ultraviolet spectroscopic study of G191-B2B]{A comprehensive near and far ultraviolet spectroscopic study of the hot DA white dwarf G191-B2B}
\author[S. P. Preval, M. A. Barstow, J. B. Holberg, \& N. J. Dickinson]{S. P. Preval$^{1}$\thanks{E-mail:
sp267@leicester.ac.uk}, M. A. Barstow$^{1}$, J. B. Holberg$^{2}$ \& N. J. Dickinson$^{1}$\\
$^{1}$Department of Physics and Astronomy, University of Leicester, University Road, Leicester, LE1 7RH\\
$^{2}$Lunar and Planetary Laboratory, Sonett Space Sciences Building, University of Arizona, Tucson, AZ 85721}
\begin{document}

\date{Accepted 2013 January 15. Received 2013 January 14; in original form 2013 January 11}

\pagerange{\pageref{firstpage}--\pageref{lastpage}} \pubyear{2002}

\maketitle

\label{firstpage}

\begin{abstract}
We present a detailed spectroscopic analysis of the hot DA white dwarf G191-B2B, using the best signal to noise, high resolution near and far UV spectrum obtained to date. This is constructed from co-added \textit{HST} STIS E140H, E230H, and \textit{FUSE} observations, covering the spectral ranges of 1150-3145\AA\, and 910-1185\AA\, respectively. With the aid of recently published atomic data, we have been able to identify previously undetected absorption features down to equivalent widths of only a few m\AA. In total, 976 absorption features have been detected to $3\sigma$ confidence or greater, with 947 of these lines now possessing an identification, the majority of which are attributed to Fe and Ni transitions. In our survey, we have also potentially identified an additional source of circumstellar material originating from Si {\sc iii}. While we confirm the presence of Ge detected by \citet{vennes05a}, we do not detect any other species. Furthermore, we have calculated updated abundances for C, N, O, Si, P, S, Fe, and Ni, while also calculating, for the first time, an NLTE abundance for Al, deriving Al {\sc iii}/H=$1.60_{-0.08}^{+0.07}\times{10}^{-7}$. Our analysis constitutes what is the most complete spectroscopic survey of any white dwarf. All observed absorption features in the \textit{FUSE} spectrum have now been identified, and relatively few remain elusive in the STIS spectrum.
\end{abstract}


\section{Introduction}
The archetype H rich white dwarf, G191-B2B (WD\,0501+527), has long been used both as a photometric standard due to its apparent brightness, and also as a spectroscopic "gold standard" when analysing other hot DA stars. This object has also been used as a flux standard at almost all wavelengths, beginning with the work of \citet{oke74a} to the absolute \textit{Hubble Space Telescope} (HST) flux scale of \citet{bohlin04a}. In Table \ref{table:factfile}, we list the basic stellar parameters of G191-B2B, where the mass, absolute magnitude ($M_\nu$) and cooling time ($t_{\mathrm{cool}}$) have been determined using the photometric tables of \citet{holberg06a,kowalski06a,tremblay11a,bergeron11a}\footnote{http://www.astro.umontreal.ca/$\sim$bergeron/CoolingModels} (hereafter the Montreal photometric tables).

The designation G191-B2B originates from the Giclas Lowell proper motion survey. While not formally identified as a white dwarf, \citet{greenstein69a} did identify it as being degenerate in nature, assigning it the spectral classification DAwk, and also designating it as a possible subdwarf. It was also listed by \citet{eggen67a} as a common proper motion pair with the K star G191-B2A, located 84" to the north. Such a proper motion pairing is now considered erroneous, as \textit{Hipparcos} has shown the two stars to have distinctly different proper motions. G191-B2B is therefore an isolated star and efforts to determine a gravitational radial velocity from both stars (\citealt{reid88a} and \citealt{bergeron95a}) are moot. A seminal high dispersion UV observation of G191-B2B by \citet{bruhweiler81a}, using the \textit{International Ultraviolet Explorer} (\textit{IUE}), revealed the surprising discovery of many highly ionised features such as C {\sc iv}, N {\sc v}, and Si {\sc iv} in a star that was thought to possess a pure H spectrum. The detection of such absorption features led to extensive studies of other white dwarf stars. A decade later, the first detections of an ionised heavy metal, Fe {\sc v}, were made by \citet{sion92a} in the photospheric spectrum of G191-B2B using the \textit{HST} Faint Object Spectrograph. This led to searches for additional heavy metals in white dwarf spectra, with the next Fe group metal, Ni, being discovered in G191-B2B and RE\,J2214-492 by \citet{holberg94a}, and Feige 24 and RE\,J0623-377 by \citet{werner94a}, using co-added \textit{IUE} high dispersion spectra. Further spectroscopic surveys were conducted, with yet more new heavy metals being discovered by \citet{vennes96a}, who made detections of the resonant transitions of P {\sc v} and S {\sc vi} in ORFEUS spectra. A later survey by \citet{bruhweiler99a} used a spectrum from the \textit{HST}, and discovered the presence of an additional component in the resonant lines of C {\sc iv} 1548.203 and 1550.777\AA\, (hereafter C {\sc iv} 1548 and 1550\AA\, respectively), that could not be attributed to either the photosphere of G191-B2B, or to the ISM. This feature was thought to be interstellar or “circumstellar” by \citet{vennes01a}, who in their survey of G191-B2B, showed that this feature was separated from the photospheric velocity by ~15 km/s. This circumstellar material was not unique to G191-B2B, and similar high ionisation features were found in seven other DA white dwarfs by \citet{bannister03a}, corroborated by \citet{dickinson12b}. The heaviest metal detected in a DA white dwarf thus far was found by \citet{vennes05a}, who made detections of resonant absorption features of Ge {\sc iv}.

The \textit{Extreme Ultraviolet Explorer} (\textit{EUVE}), and the \textit{Joint Astrophysical Plasma-dynamic Experiment} (\textit{J-PEX}) \citep{bannister99a} have observed G191-B2B extensively, covering 70-770\AA\, and 170-290\AA\, respectively. \textit{EUVE} observations of the star showed that the predicted flux far exceeded that observed \citep{kahn84a}, meaning that model atmospheres for some effective temperature ($T_{\mathrm{eff}}$), gravity (log $g$), and composition could not simultaneously match the EUV, UV, and optical spectra. This was accomplished by \citet{lanz96a}, who showed that including additional opacity in model atmosphere calculations reduced the predicted flux to the correct level. However, this came at the cost of including additional He {\sc ii} opacity in the photosphere in order to maintain good agreement below the He {\sc ii} Lyman limit. An alternative was explored by \citet{barstow98a}, whereby they invoked a stratified H+He and homogeneous heavy metal atmosphere. While the model could also reproduce the EUV spectrum, it predicted He {\sc ii} absorption features that descended far deeper than that observed. A substantial piece of work was done by \citet{dreizler99a}, who performed self consistent calculations including the effects of radiative levitation and gravitation settling for G191-B2B. By depleting the Fe and Ni abundance at the surface and having it increase at greater depths, they were able to reproduce the EUV spectrum without including other sources of opacity, mass loss, or accretion. A similar atmospheric configuration was used by \citet{barstow99a}, who stratified Fe, with increasing abundance with greater depth. They showed that this stratified configuration was preferred statistically over a homogeneous Fe distribution. Observations with \textit{J-PEX} yielded further discoveries in the EUV spectrum of G191-B2B. \citet{cruddace02a} performed a spectroscopic analysis of the star using \textit{J-PEX} data, and made detections of the He {\sc ii} Lyman series. This would be significant if it was found to be interstellar, as this would imply an unusual ionization fraction. It was also proposed that this He may be associated with the additional absorption component observed in the C {\sc iv} doublet, discussed by \citet{bruhweiler99a} and \citet{vennes01a}. A further spectroscopic survey using \textit{J-PEX} was conducted by \citet{barstow05a}. They found that by splitting the He {\sc ii} features into two components, one for the Local Interstellar Cloud, and the other as another interstellar feature, the He ionisation fraction agreed with values obtained from other lines of sight.

Observations of the H Lyman/Balmer line series of a white dwarf also allows information on its physical parameters to be inferred. Pioneered by \citet{holberg85a}, a grid of theoretical spectra with differing values of $T_{\mathrm{eff}}$ and log $g$ can be used to fit either the H Lyman or Balmer absorption profiles to that of an observed spectrum, as such profiles are very sensitive to changes in these parameters. Such a method, however, appears to have limitations for white dwarfs whose $T_{\mathrm{eff}}>$40,000K. Dubbed the "Lyman-Balmer line problem", measurements of $T_{\mathrm{eff}}$ made in DA white dwarfs using either the Lyman or Balmer line series yield differing values, some by a few 1,000K, and some even by 10,000K, getting larger for increasing values of $T_{\mathrm{eff}}$ \citep{barstow03a}. No similar appreciable effect is seen on the measurement of log $g$. A similar, more severe effect is seen in DAO stars \citep{good04a}. A study by \citet{lajoie07a} considered many different causes of this problem, ranging from atmospheric composition to unresolved binaries. One such cause was concluded to be due to some form of wavelength dependent extinction, however, the exact relation between the two was left open to debate. It was also suggested that atmospheric composition may play a part in causing the discrepancy, but results were inconclusive. It is not unreasonable to consider such an idea, however, as metal line blanketing dramatically effects a hot DA SED. \citet{barstow98a} have shown that using a model grid with a pure H atmosphere and a grid with a line blanketed heavy metal atmosphere yield $T_{\mathrm{eff}}$ measurements again differing by 1000s of K. 

\citet{barstow03a} also considered the Lyman-Balmer line problem, postulating that the temperature discrepancy may be due to inadequacies in the input atomic physics in generating the model atmospheres. Using G191-B2B and RE\,J2214-492, they calculated two model grids with varying $T_{\mathrm{eff}}$ and log $g$, but with 0.1 and 10 times the nominal abundances of these stars. They then measured $T_{\mathrm{eff}}$ and log $g$ using the Lyman and Balmer line series, finding that while the temperature difference decreased by 20-30\%, a statistically significant discrepancy still remained. 

An accurate knowledge of a star's photospheric composition is, therefore, paramount in accurately constraining $T_{\mathrm{eff}}$ and log $g$ in model atmosphere calculations. Determination of these parameters can provide useful insights into the white dwarf's origin and evolution. For example, the cooling time of a degenerate object has a one to one correspondence with $T_{\mathrm{eff}}$, and hence gives a monotonic cooling age. Knowledge of log $g$ yields information on the mass of the white dwarf, and along with evolutionary models such as those from \citet{wood98a} or the Montreal Photometric Tables, an estimate of the radius of the star. 

Calculations by \citet{chayer95a} for various values of $T_{\mathrm{eff}}$ and log $g$ describe the predicted variation of metal abundances due to radiative levitation. However, the observational results do not agree very well with the predictions. In the case of G191-B2B for example, \citet{barstow03b} reported the Fe and Ni abundances to be $3.30_{-1.20}^{+3.10}\times{10^{-6}}$ and $2.40_{-0.24}^{+0.84}\times{10^{-7}}$ respectively, differing roughly by an order of magnitude. However, \citet{chayer95a} predict that these abundances should share a similar value. This disagreement is also present in other white dwarfs in the sample of \citet{barstow03b}. It should be noted, however, that \citet{chayer94a} reported significant differences in predicted atmospheric abundances of Fe and Ni dependent on the number of transitions included in their calculations. It is this variation in results that suggests that poor agreement between theory and observation may be dependent on the number of opacities included in these calculations. 

\begin{table}
\centering
\caption{A summary of the physical parameters of G191-B2B. The velocities $v_{\mathrm{LIC}}$ and $v_{\mathrm{Hyades}}$ are calculated along the line of sight to the star.}
\begin{tabular}[H]{@{}lll}
\hline
Parameter              & Value            & Reference \\
\hline
V                      & $11.727\pm{0.016}$    & \protect\citet{holberg06a} \\
$M_{v}$                & $8.280\pm{0.164}$     & This work \\
$T_{\mathrm{eff}}$ (K) & $52,500\pm{900}$      & \protect\citet{barstow03b}\\
Log $g$                & $7.53\pm{0.09}$       & \protect\citet{barstow03b}\\
Mass ($M_{\odot}$)     & $0.52\pm{0.035}$     & This work \\
Radius ($R_{\odot}$)   & $0.0204\pm{0.0014}$ & This work \\
Distance (Pc)          & $48.9\pm{3.7}$    & This work \\
$t_{\mathrm{cool}}$ (Myr) & $1.50\pm{0.08}$   & This work \\
$v_{\mathrm{phot}}$ (km s$^{-1}$) & $23.8\pm{0.03}$       & This work \\
$v_{\mathrm{LIC}}$ (km s$^{-1}$) & $19.4\pm{0.03}$       & This work \\
$v_{\mathrm{Hyades}}$ (km s$^{-1}$) & $8.64\pm{0.03}$       & This work \\
\hline
\end{tabular}
\label{table:factfile}
\end{table}

In this paper we present and analyse a unique spectrum of G191-B2B with unprecedented signal to noise (S/N), over the wavelength range 910-3145\AA. The spectrum is constructed using co-added \textit{Far Ultraviolet Spectroscopic Explorer} (\textit{FUSE}) LWRS (30$\times$30", low resolution), MDRS (4.0$\times$20", medium resolution), and HIRS (1.21$\times$20", high resolution) spectra, and co-added Space Telescope Imaging Spectrometer (STIS) E140H and E230H spectra. The resolution of our data sets range from 25,000 for the \textit{FUSE} data to 144,000 for the \textit{HST} spectra. The S/N exceeds 100 in many regions. We begin by describing the observational data used, and how co-addition of several data sets has allowed access to previously undetectable absorption lines. In section 3, we discuss the new atomic data releases provided by the Kurucz\footnote{http://kurucz.harvard.edu} (\citealt{kurucz92a,kurucz06a,kurucz11a}, hereafter Kurucz) and Kentucky\footnote{http://www.pa.uky.edu/$\sim$peter/newpage/} (hereafter Kentucky) databases and the effect this has had on our ability to identify new lines. A comprehensive table of identifications has been included in Appendix B. We calculate the atmospheric abundances of C, N, O, Al, Si, P, S, Fe, and Ni. We discuss a new potential circumstellar identification, and the abundance pattern of the white dwarf. The potential significance of including additional opacities into model atmosphere calculations is also discussed, providing a tentative solution to the Lyman-Balmer line problem. 
In summary, this data set is a unique and invaluable record of the archetype white dwarf G191-B2B. It is the most complete, and highest S/N NUV and FUV spectrum available for any white dwarf observed. It will serve as a template for studies of other hot white dwarfs, as a test bed for model atmosphere calculations, and also for the improvement of atomic databases. 

\section{Observational Data}
Three detailed coadded spectra were used to analyse the NUV/FUV flux distribution of G191-B2B, spanning 910 to 3145\AA. All data used in the coadded spectra are hosted on the Mikulski Archive for Space Telescopes\footnote{http://archive.stsci.edu/} (MAST). Using LWRS, MDRS, and HIRS exposures observed by \textit{FUSE}, \citet{barstow10a} constructed a coadded spectrum spanning 910-1185\AA\, with an exceptional S/N ratio. Observations by STIS aboard the \textit{HST} covered the remainder of the spectrum. We made use of as many high resolution ($R\approx{144,000}$) observations with the echelle gratings E140H (centroid wavelength 1400\AA) and E230H (centroid wavelength 2300\AA) as possible, and produced a coadded spectrum spanning 1160-1680\AA\, and 1625-3145\AA\, respectively. 

\subsection{\textit{FUSE}}
\begin{figure*}
\begin{centering}
\includegraphics[width=120mm]{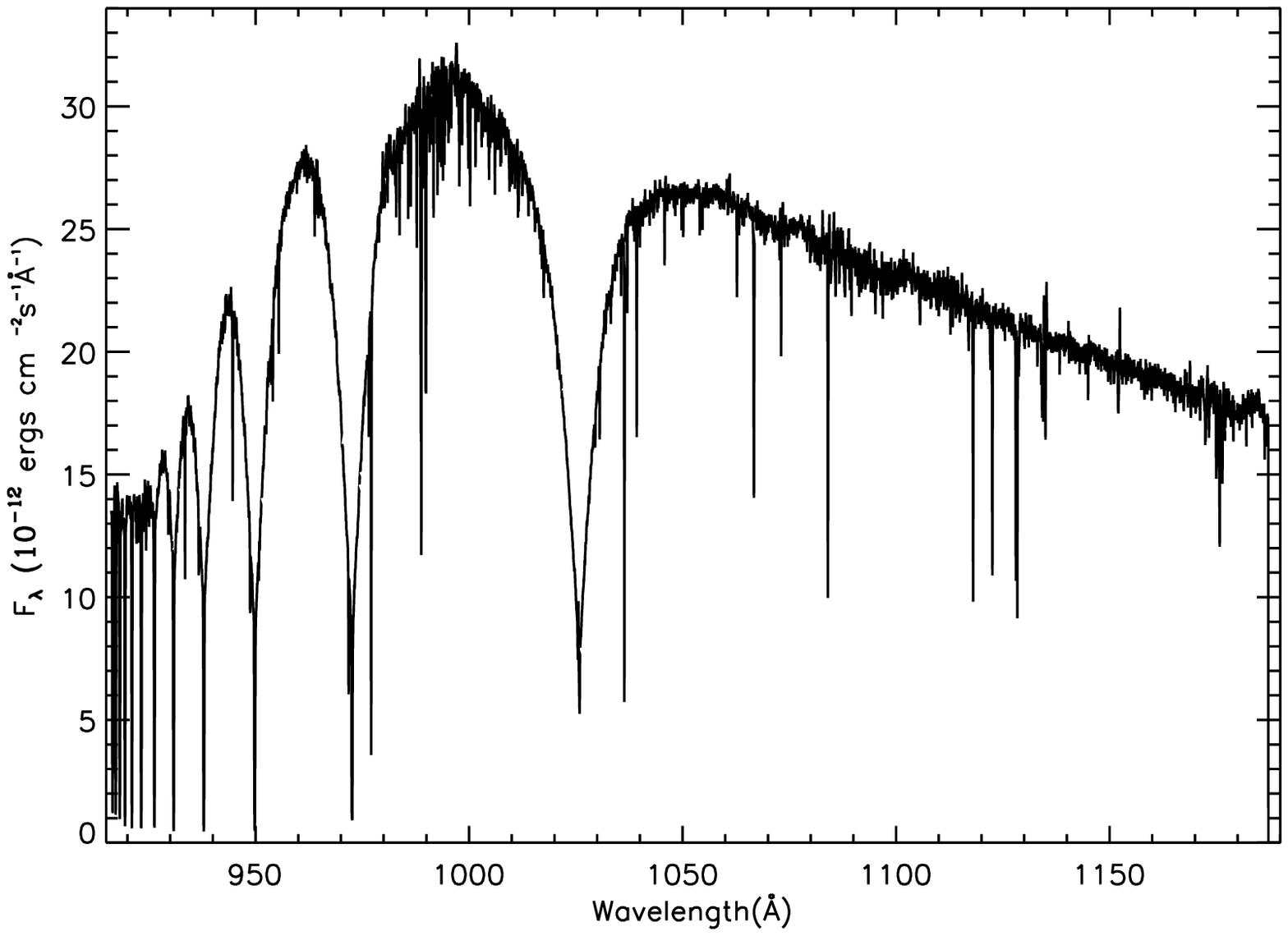}
\caption{Co-added spectrum of G191-B2B using 48 \textit{FUSE} observations from LWRS, MDRS, and HIRS apertures.}
\label{fig:fuseplot}
\end{centering}
\end{figure*}

\textit{FUSE} was launched in 1999, providing coverage from 910\AA\, to 1185\AA, making the Lyman series accessible to Ly $\beta$. While the satellite has been described many times (e.g. \citealt{moos00a}), we provide a brief summary of the hardware here. \textit{FUSE} utilised a Rowland Circle design with four separate channels or coaligned optical paths. The satellite employed two detectors, each of which had two independent segments, SiC and LiF, upon which the spectra from the four channels are recorded. \textit{FUSE} has three different aperture configurations, LWRS, MDRS and HIRS. \textit{FUSE} also has a pinhole aperture (RFPT), which was mainly used for calibration. To minimise the possibility that the target's light did not fall on the detector, observations were conducted primarily using the LWRS aperture, with a spectral resolution of 15,000-20,000 for early observations, and 23,000 for later ones, where the mirror focusing had been adjusted \citep{sahnow00a}. Observations were also taken using both the TIMETAG (TTAG) and HISTOGRAM (HIST) modes. All \textit{FUSE} data used here were reduced and processed using version 3.2 of {\sc calfuse} \citep{dixon07a}. The data comprises exposures from each of the detector/segment combinations (eight in total). We used the FUSE spectrum of G191-B2B from \citet{barstow10a}, constructed with 48 observations listed in Table \ref{table:exposurelist}, and plotted in Figure \ref{fig:fuseplot}. The different observations were coadded according to the process described by \citet{barstow03a}. Prior to coaddition, all spectra were rebinned to a common wavelength spacing to account for the difference in resolution between apertures. A consequence of coadding exposures from the different slits are discontinuities in the flux. It is for this reason we applied a correcting factor of 1.08 shortward of 1089\AA\, to ensure continuity between the FUSE and STIS flux distributions.

\begin{table*}
\centering
\caption{The \textit{FUSE} datasets obtained from MAST.}
\begin{tabular}[H]{@{}ccccc}
\hline
Observation ID & Number of exposures & Start time & Exposure time (s) & Aperture \\
\hline
M1010201000 & 8 & 13/10/99 01:25 & 4164 & LWRS \\
M1030501000 & 1 & 12/11/99 07:35 & 266 & MDRS \\
M1030502000 & 1 & 20/11/99 07:22 & 900 & MDRS \\
M1030401000 & 1 & 20/11/99 09:02 & 1298 & HIRS \\
M1030603000 & 5 & 20/11/99 10:43 & 3664 & LWRS \\
M1030503000 & 3 & 21/11/99 06:43 & 1709 & MDRS \\
M1030504000 & 4 & 21/11/99 10:03 & 3212 & MDRS \\
M1030602000 & 5 & 21/11/99 11:39 & 2812 & LWRS \\
S3070101000 & 32 & 14/01/00 09:40 & 15456 & LWRS \\
M1010202000 & 7 & 17/02/00 06:10 & 3450 & LWRS \\
M1030604000 & 1 & 09/01/01 09:02 & 503 & LWRS \\
M1030506000 & 1 & 09/01/01 09:26 & 503 & MDRS \\
M1030605000 & 1 & 10/01/01 13:20 & 503 & LWRS \\
M1030507000 & 1 & 10/01/01 13:45 & 503 & MDRS \\
M1030403000 & 2 & 10/01/01 15:08 & 483 & HIRS \\
M1030606000 & 5 & 23/01/01 06:08 & 2190 & LWRS \\
M1030508000 & 5 & 23/01/01 07:55 & 2418 & MDRS \\
M1030404000 & 5 & 23/01/01 11:18 & 1853 & HIRS \\
M1030607000 & 5 & 25/01/01 04:46 & 1926 & LWRS \\
M1030509000 & 5 & 25/01/01 06:33 & 2417 & MDRS \\
M1030405000 & 5 & 25/01/01 09:53 & 2419 & HIRS \\
M1030608000 & 5 & 28/09/01 13:50 & 2728 & LWRS \\
M1030510000 & 4 & 28/09/01 15:35 & 1910 & MDRS \\
M1030406000 & 5 & 28/09/01 17:15 & 1932 & HIRS \\
M1030609000 & 5 & 21/11/01 09:54 & 2703 & LWRS \\
M1030511000 & 4 & 21/11/01 11:39 & 1910 & MDRS \\
M1030407000 & 5 & 21/11/01 13:19 & 1932 & HIRS \\
M1030610000 & 16 & 17/02/02 07:27 & 8639 & LWRS \\
M1030512000 & 11 & 17/02/02 12:34 & 4757 & MDRS \\
M1030408000 & 5 & 17/02/02 17:43 & 1932 & HIRS \\
M1030611000 & 8 & 23/02/02 02:05 & 3645 & LWRS \\
M1030513000 & 5 & 23/02/02 06:43 & 1797 & MDRS \\
M1030409000 & 5 & 23/02/02 08:23 & 1921 & HIRS \\
M1030612000 & 14 & 25/02/02 02:17 & 7004 & LWRS \\
M1030514000 & 4 & 25/02/02 06:59 & 1617 & MDRS \\
M1030613000 & 5 & 03/12/02 21:00 & 2358 & LWRS \\
M1030614000 & 3 & 06/12/02 02:30 & 702 & LWRS \\
M1030515000 & 4 & 06/12/02 05:16 & 2002 & MDRS \\
M1052001000 & 16 & 07/12/02 21:46 & 7061 & LWRS \\
M1030615000 & 4 & 08/12/02 22:36 & 1895 & LWRS \\
M1030516000 & 4 & 09/12/02 00:29 & 1911 & MDRS \\
M1030412000 & 4 & 09/12/02 03:53 & 1880 & HIRS \\
M1030616000 & 4 & 05/02/03 19:14 & 1980 & LWRS \\
M1030517000 & 4 & 05/02/03 21:07 & 1910 & MDRS \\
M1030413000 & 4 & 06/02/03 00:36 & 1932 & HIRS \\
M1030617000 & 8 & 23/11/03 20:16 & 4121 & LWRS \\
M1030519000 & 4 & 25/01/04 21:31 & 1887 & MDRS \\
M1030415000 & 4 & 26/01/04 00:51 & 1902 & HIRS \\
\hline
\end{tabular}
\label{table:exposurelist}
\end{table*}

\subsection{STIS}
G191-B2B has been extensively observed as a calibration standard by \textit{HST}. In particular, observations of this star were conducted as part of Cycle 8 STIS calibration programs 8067, 8421 and 8915, which were designed to provide flux calibrations at the 1\% level for all E140H and E230H primary and secondary echelle grating modes with a strong stellar continuum source. After the 2009 repair mission STS-125, an additional calibration program 11866 was proposed in Cycle 17 in order to evaluate the post-repair echelle blaze dependence on MSM position. The data were obtained in the ACCUM mode in four periods; 17th December 1998, 16th to 19th March 2000, 17th to 19th September 2001, and 28th November 2009 to 6th January 2010. Standard target acquisition procedures were used to acquire the source and centre it within the 0.2$\times$0.2" slit. The wavelength range 1140-3145\AA\, was covered by using all nine STIS primary grating settings and 28 secondary grating settings (see Chapter 11 of \citealt{stishand12}). 

\begin{figure*}
\begin{centering}
\includegraphics[width=120mm,clip=true,trim=0mm 0mm 0mm 0mm]{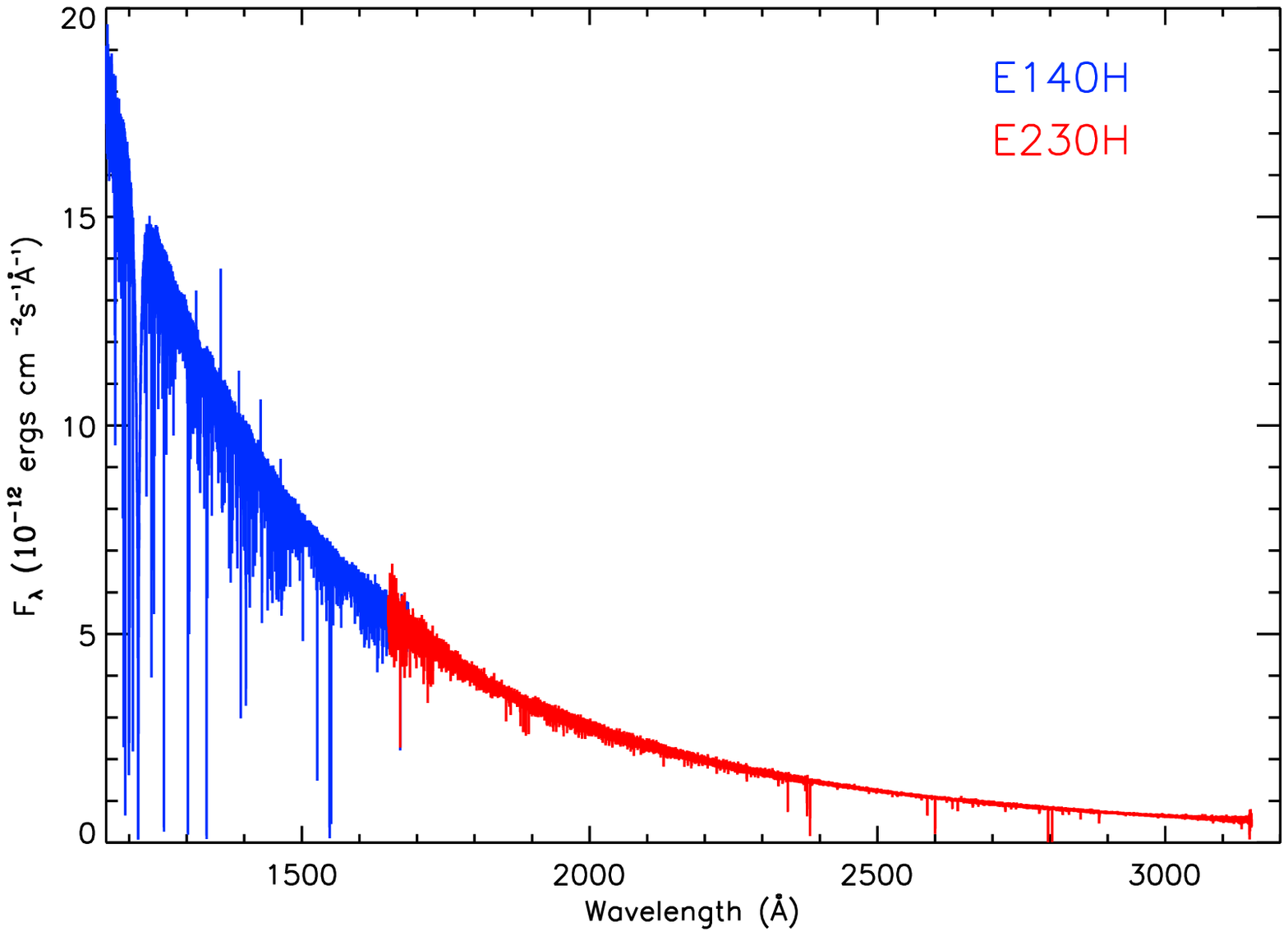}
\caption{The final coadded spectra for the E140H (blue in online copy) and the E230H spectra (red in online copy). Most of the features falling below the continuum are identified absorption features.}
\label{fig:1stis}
\end{centering}
\end{figure*}

\begin{figure*}
\begin{centering}
\includegraphics[width=120mm,clip=true,trim=0mm 0mm 0mm 0mm]{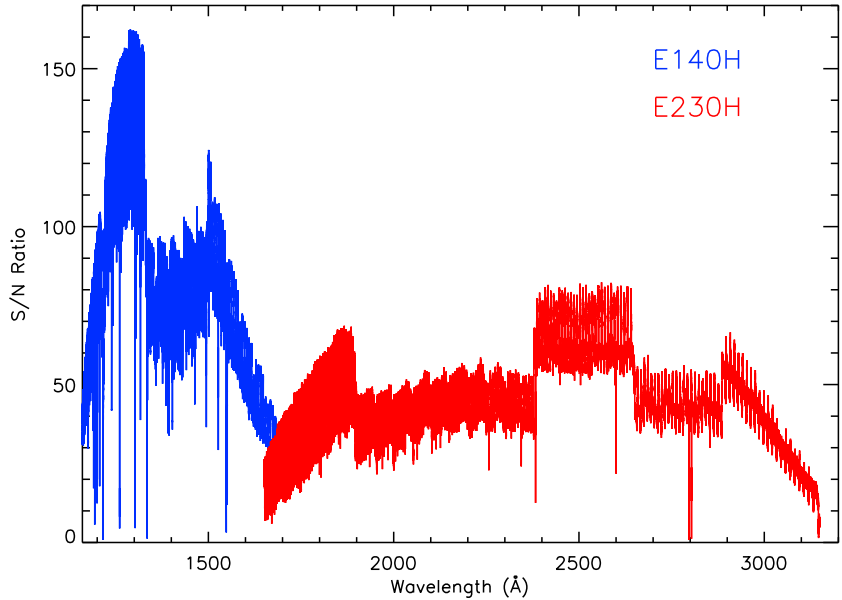}
\caption{The dependence of signal to noise with wavelength. The wavelength ranges (and colour scheme in the online version) are the same as for Figure \protect\ref{fig:1stis}.}
\label{fig:2stis}
\end{centering}
\end{figure*}

We examined all of the available E140H and E230H datasets (39 and 77 respectively) from the MAST website in order to check for discontinuities and errors in the observations. We found that 32 E140H and 66 E230H observations were free of such problems, and were hence included in the final coadded product, with total exposure times of 53318 and 77743s respectively. We summarise the individual STIS spectra in Table \ref{table:2stis}. After the extraction of the echelle orders, including ripple correction, the spectra were interpolated on to a single linear wavelength scale prior to exposure time weighted coaddition. The result was two single continuous spectra as shown in Figure \ref{fig:1stis}. The distribution of S/N as a function of wavelength is shown in Figure \ref{fig:2stis}. 

\begin{table*}
\centering
\caption{A list of exposures used in the STIS spectrum coaddition. Where $\lambda_{c}$ is the central wavelength of the exposure, P is the primary setting, and S is the secondary setting. }
\begin{tabular}[H]{@{}llccccc}
\hline
Observation ID  &  Date  &  Prog. ID  &  Grating  &  $\lambda_{c}$ (\AA)  &  Setting  &  Exposure time (s)  \\
\hline
O57U01020 & 17/12/1998 08:17 & 8067 & E140H & 1416 & P & 2040 \\
O57U01030 & 17/12/1998 09:34 & 8067 & E140H & 1234 & P & 2789 \\
O57U01040 & 17/12/1998 11:14 & 8067 & E140H & 1598 & P & 2703 \\
O5I010010 & 16/03/2000 23:31 & 8421 & E140H & 1234 & P & 2279 \\
O5I010020 & 17/03/2000 00:52 & 8421 & E140H & 1234 & P & 3000 \\
O5I010030 & 17/03/2000 02:29 & 8421 & E140H & 1234 & P & 3000 \\
O5I011010 & 17/03/2000 04:21 & 8421 & E140H & 1598 & P & 2284 \\
O5I011020 & 17/03/2000 05:42 & 8421 & E140H & 1598 & P & 3000 \\
O5I011030 & 17/03/2000 07:18 & 8421 & E140H & 1598 & P & 3000 \\
O5I014010 & 18/03/2000 02:52 & 8421 & E230H & 2513 & P & 2304 \\
O5I014020 & 18/03/2000 04:14 & 8421 & E230H & 2513 & P & 3000 \\
O5I014030 & 18/03/2000 05:50 & 8421 & E230H & 2513 & P & 3000 \\
O5I015010 & 18/03/2000 22:10 & 8421 & E230H & 3012 & P & 2304 \\
O5I015020 & 18/03/2000 23:32 & 8421 & E230H & 3012 & P & 3000 \\
O5I015030 & 19/03/2000 01:09 & 8421 & E230H & 3012 & P & 3000 \\
O5I013010 & 19/03/2000 03:00 & 8421 & E230H & 1763 & P & 2304 \\
O5I013020 & 19/03/2000 04:22 & 8421 & E230H & 1763 & P & 3000 \\
O5I013030 & 19/03/2000 05:59 & 8421 & E230H & 1763 & P & 3000 \\
O6HB40080 & 12/09/2001 23:09 & 8915 & E230H & 2413 & S & 774 \\
O6HB40090 & 13/09/2001 00:08 & 8915 & E230H & 3012 & P & 2228 \\
O6HB10010 & 17/09/2001 13:49 & 8915 & E140H & 1234 & P & 867 \\
O6HB10020 & 17/09/2001 14:05 & 8915 & E140H & 1234 & P & 867 \\
O6HB10040 & 17/09/2001 14:54 & 8915 & E140H & 1271 & S & 640 \\
O6HB10050 & 17/09/2001 15:11 & 8915 & E140H & 1307 & S & 654 \\
O6HB10080 & 17/09/2001 16:30 & 8915 & E140H & 1380 & S & 719 \\
O6HB10090 & 17/09/2001 16:48 & 8915 & E140H & 1416 & P & 851 \\
O6HB100A0 & 17/09/2001 17:09 & 8915 & E140H & 1453 & S & 809 \\
O6HB100B0 & 17/09/2001 18:07 & 8915 & E140H & 1453 & S & 229 \\
O6HB100C0 & 17/09/2001 18:17 & 8915 & E140H & 1489 & S & 1263 \\
O6HB100D0 & 17/09/2001 18:44 & 8915 & E140H & 1526 & S & 887 \\
O6HB100E0 & 17/09/2001 19:43 & 8915 & E140H & 1526 & S & 749 \\
O6HB100F0 & 17/09/2001 20:02 & 8915 & E140H & 1562 & S & 1996 \\
O6HB20010 & 18/09/2001 15:32 & 8915 & E230H & 1763 & P & 1314 \\
O6HB20020 & 18/09/2001 16:00 & 8915 & E230H & 1813 & S & 654 \\
O6HB20030 & 18/09/2001 16:35 & 8915 & E230H & 1813 & S & 455 \\
O6HB20040 & 18/09/2001 16:49 & 8915 & E230H & 1863 & S & 997 \\
O6HB20050 & 18/09/2001 17:12 & 8915 & E230H & 1913 & S & 907 \\
O6HB20060 & 18/09/2001 18:11 & 8915 & E230H & 1963 & S & 871 \\
O6HB20070 & 18/09/2001 18:32 & 8915 & E230H & 2013 & P & 808 \\
O6HB20080 & 18/09/2001 18:51 & 8915 & E230H & 2063 & S & 718 \\
O6HB20090 & 18/09/2001 19:47 & 8915 & E230H & 2113 & S & 679 \\
O6HB200A0 & 18/09/2001 20:05 & 8915 & E230H & 2163 & S & 640 \\
O6HB200B0 & 18/09/2001 20:21 & 8915 & E230H & 2213 & S & 620 \\
O6HB200C0 & 18/09/2001 20:38 & 8915 & E230H & 2263 & P & 101 \\
O6HB200D0 & 18/09/2001 21:24 & 8915 & E230H & 2263 & P & 609.6 \\
O6HB200E0 & 18/09/2001 21:40 & 8915 & E230H & 2313 & S & 734.3 \\
O6HB200F0 & 18/09/2001 21:59 & 8915 & E230H & 2363 & S & 748.7 \\
O6HB30010 & 19/09/2001 15:39 & 8915 & E230H & 2463 & S & 668 \\
O6HB30020 & 19/09/2001 15:56 & 8915 & E230H & 2513 & P & 696 \\
O6HB30030 & 19/09/2001 16:13 & 8915 & E230H & 2563 & S & 247 \\
O6HB30040 & 19/09/2001 16:39 & 8915 & E230H & 2563 & S & 484 \\
O6HB30050 & 19/09/2001 16:54 & 8915 & E230H & 2613 & S & 769 \\
O6HB30060 & 19/09/2001 17:13 & 8915 & E230H & 2663 & S & 992.3 \\
O6HB30070 & 19/09/2001 18:16 & 8915 & E230H & 2713 & S & 900 \\
O6HB30080 & 19/09/2001 18:37 & 8915 & E230H & 2762 & P & 978 \\
O6HB30090 & 19/09/2001 18:59 & 8915 & E230H & 2812 & S & 519 \\
O6HB300A0 & 19/09/2001 19:52 & 8915 & E230H & 2812 & S & 578 \\
O6HB300B0 & 19/09/2001 20:08 & 8915 & E230H & 2862 & S & 1232 \\
O6HB300C0 & 19/09/2001 20:35 & 8915 & E230H & 2912 & S & 549 \\
O6HB300D0 & 19/09/2001 21:28 & 8915 & E230H & 2912 & S & 873 \\
\hline
\end{tabular}
\label{table:2stis}
\end{table*}

\begin{table*}
\centering
\contcaption{}
\begin{tabular}[H]{@{}llccccc}
\hline
Observation ID  &  Date  &  Prog. ID  &  Grating  &  $\lambda_{c}$ (\AA)  &  Setting  &  Exposure time (s)  \\
\hline
O6HB300E0 & 19/09/2001 21:49 & 8915 & E230H & 2962 & S & 1862 \\
OBB002010 & 28/11/2009 08:36 & 11866 & E230H & 1863 & S & 1000 \\
OBB002020 & 28/11/2009 08:59 & 11866 & E230H & 1963 & S & 870 \\
OBB002030 & 28/11/2009 09:55 & 11866 & E230H & 1913 & S & 920 \\
OBB002040 & 28/11/2009 10:16 & 11866 & E230H & 2013 & P & 810 \\
OBB002050 & 28/11/2009 10:36 & 11866 & E230H & 2063 & S & 740 \\
OBB002060 & 28/11/2009 11:31 & 11866 & E230H & 2263 & P & 800 \\
OBB002070 & 28/11/2009 11:50 & 11866 & E230H & 2113 & S & 850 \\
OBB002080 & 28/11/2009 12:10 & 11866 & E230H & 2163 & S & 800 \\
OBB002090 & 28/11/2009 13:07 & 11866 & E230H & 1763 & P & 1800 \\
OBB0020A0 & 28/11/2009 13:43 & 11866 & E230H & 2213 & S & 1000 \\
OBB0020B0 & 28/11/2009 14:43 & 11866 & E230H & 1813 & S & 1160 \\
OBB0020C0 & 28/11/2009 15:08 & 11866 & E230H & 2313 & S & 650 \\
OBB0020D0 & 28/11/2009 15:25 & 11866 & E230H & 2363 & S & 650 \\
OBB004080 & 29/11/2009 12:12 & 11866 & E230H & 2413 & S & 645 \\
OBB004090 & 29/11/2009 13:05 & 11866 & E230H & 3012 & P & 2192 \\
OBB001010 & 30/11/2009 06:58 & 11866 & E140H & 1271 & S & 696 \\
OBB001020 & 30/11/2009 07:15 & 11866 & E140H & 1453 & S & 1038 \\
OBB001030 & 30/11/2009 08:16 & 11866 & E140H & 1380 & S & 752 \\
OBB001040 & 30/11/2009 08:34 & 11866 & E140H & 1234 & P & 867 \\
OBB001050 & 30/11/2009 08:54 & 11866 & E140H & 1416 & P & 851 \\
OBB001070 & 30/11/2009 10:09 & 11866 & E140H & 1526 & S & 2100 \\
OBB001080 & 30/11/2009 11:27 & 11866 & E140H & 1562 & S & 2134 \\
OBB001090 & 30/11/2009 12:09 & 11866 & E140H & 1307 & S & 654 \\
OBB0010A0 & 30/11/2009 13:03 & 11866 & E140H & 1489 & S & 1200 \\
OBB005010 & 01/12/2009 05:12 & 11866 & E140H & 1234 & P & 2200 \\
OBB005020 & 01/12/2009 06:38 & 11866 & E140H & 1234 & P & 6200 \\
OBB053010 & 06/01/2010 13:30 & 11866 & E230H & 2563 & S & 900 \\
OBB053020 & 06/01/2010 13:51 & 11866 & E230H & 2613 & S & 950 \\
OBB053030 & 06/01/2010 14:43 & 11866 & E230H & 2663 & S & 830 \\
OBB053040 & 06/01/2010 15:03 & 11866 & E230H & 2463 & S & 670 \\
OBB053050 & 06/01/2010 15:20 & 11866 & E230H & 2713 & S & 900 \\
OBB053060 & 06/01/2010 16:19 & 11866 & E230H & 2762 & P & 1197 \\
OBB053070 & 06/01/2010 16:45 & 11866 & E230H & 2862 & S & 1647 \\
OBB053080 & 06/01/2010 17:55 & 11866 & E230H & 2513 & P & 1000 \\
OBB053090 & 06/01/2010 18:18 & 11866 & E230H & 2912 & S & 1800 \\
OBB0530A0 & 06/01/2010 19:31 & 11866 & E230H & 2812 & S & 1097 \\
OBB0530B0 & 06/01/2010 19:55 & 11866 & E230H & 2962 & S & 1747 \\
\hline
\end{tabular}
\label{table:2stis}
\end{table*}

\section{Line survey}

\subsection{Atomic data}
\begin{figure*}
\begin{centering}
\includegraphics[width=160mm,clip=true,trim=0mm 0mm 0mm 0mm]{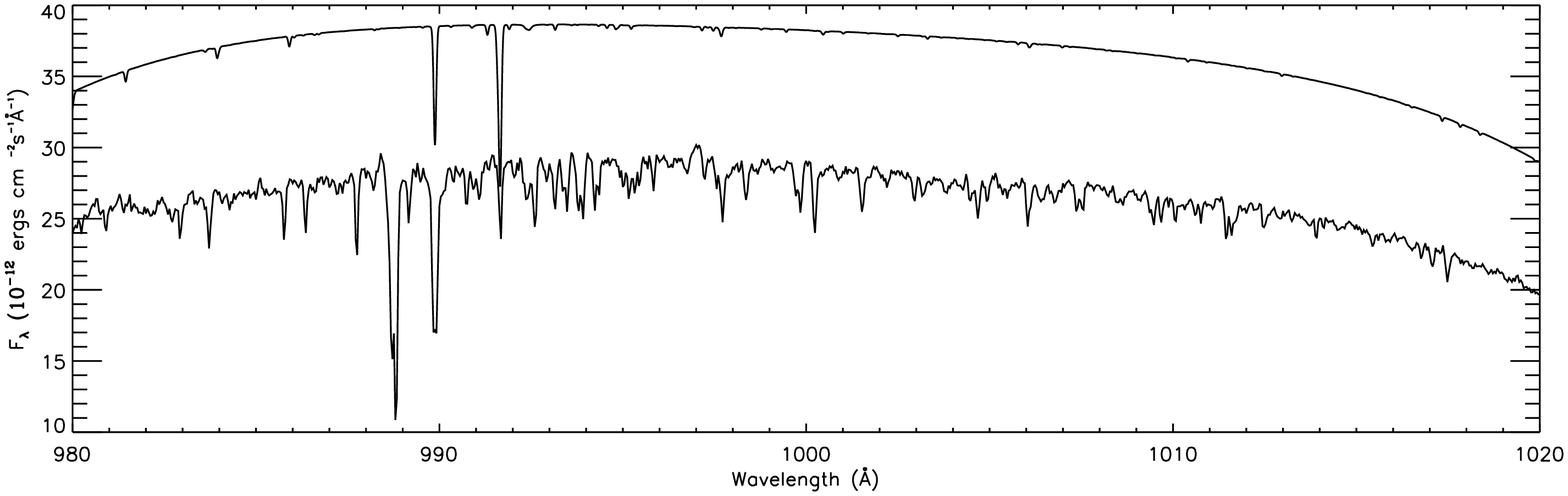}
\includegraphics[width=160mm,clip=true,trim=0mm 0mm 0mm 0mm]{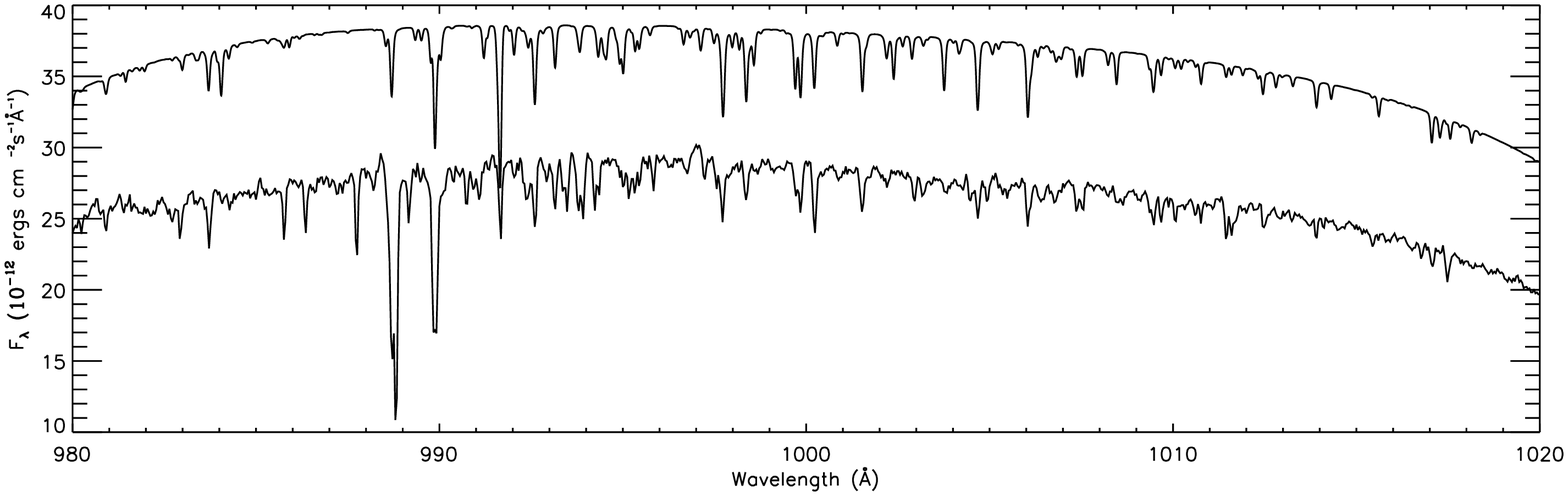}
\caption{A comparison of the predicted synthetic (upper line) spectrum to the observed spectrum (bottom line). In the top plot, the spectrum has been synthesised using only the 1992 Kurucz data release and some lines from NIST (described in text), whereas in the bottom plot, the spectrum has been synthesised using both the 2011 Kurucz and Kentucky data releases. The synthetic spectrum has been offset for clarity.}
\label{fig:oldnewcom}
\end{centering}
\end{figure*}

We combined data from the Kurucz and Kentucky atomic databases to compile as complete a line list as possible. The Kurucz database has been updated several times, with major updates occurring in 1992, 2006, and 2011 \citep{kurucz92a,kurucz06a,kurucz11a}. Table \ref{table:lines} shows how the number of transitions available for Fe {\sc iv}-{\sc vii} and Ni {\sc iv}-{\sc vii} have increased from the 1992 to the 2011 data releases; an order of magnitude increase in the number of transitions is seen for each ion. Figure \ref{fig:oldnewcom} illustrates the improvement made in reproducing the 980-1020\AA\, spectral region of G191-B2B using the 1992 Kurucz data release and some lines from the National Institute of Standards and Technology\footnote{See http://nova.astro.umd.edu/Synspec49/data/ for more information on this line list.} (NIST), and our combined line list, synthesised using the model atmosphere of \cite{barstow03b}. Some transitions in the Kentucky database lacked oscillator strengths ($f$-values). For Figure \ref{fig:oldnewcom}, we set the missing $f$-values to $1.00\times{10}^{-6}$ for illustrative purposes. We choose this value as weak Fe/Ni transitions have oscillator strengths $\sim{10^{-6}}$. The Kurucz and Kentucky databases often had records of the same transition. For the purposes of identification, we used the data from the Kentucky database, as this supplies errors on the laboratory wavelengths of transitions. To identify resonant transitions, we used the line list of \citet{verner94a} (hereafter V94), and found the error on the wavelength by cross-correlating this line list with the Kentucky database.

\begin{table}
\centering
\caption{The number of lines present in the Kurucz linelist in 1992 and 2011 for Fe and Ni {\sc iv}-{\sc vii}.}
\begin{tabular}[H]{@{}lrr}
\hline
Ion & No of lines 1992 & No of lines 2011 \\
\hline
Fe {\sc iv} & 1,776,984 & 14,617,228 \\
Fe {\sc v} & 1,008,385 & 7,785,320 \\
Fe {\sc vi} & 475,750 & 9,072,714 \\
Fe {\sc vii} & 90,250 & 2,916,992 \\
Ni {\sc iv} & 1,918,070 & 15,152,636 \\
Ni {\sc v} & 1,971,819 & 15,622,452 \\
Ni {\sc vi} & 2,211,919 & 17,971,672 \\
Ni {\sc vii} & 967,466 & 28,328,012 \\
Total & 10,420,643 & 111,467,026 \\
\hline
\end{tabular}
\label{table:lines}
\end{table}

\subsection{Absorption feature parameterisation}
For the purposes of identification, we need to extract the basic parameters of each absorption feature, such as wavelength centroid, velocity, and equivalent width. To parameterise the absorption feature as accurately as possible, we fit a Gaussian and Lorentzian profile (see Appendix A for the exact parameterisation) to the feature with a chi squared ($\chi^2$) minimisation technique, using the IDL routine {\sc mpfit} \citep{markwardt09a}. The fit that achieved the lowest $\chi^2$ was assumed to be the best fit. Hence we used the calculated parameters from this fit. To differentiate between signal and noise, we adopted a $3\sigma$ threshold. In the cases where absorption features appeared to be blended, we used a double Gaussian absorption profile (see also Appendix A).

\subsection{Detections and identification}
We detected 976 absorption features, successfully identifying 947 of them. Lines that could not be identified are listed in Table \ref{table:unidentified}. Several measured lines in our survey were found to be the result of multiple blended features. Therefore, the velocity of each individual absorber was included in calculating the weighted velocity of each component. To calculate the uncertainty on the velocity measurements, the uncertainties in the measured wavelength ($\lambda_{\mathrm{tot}}$) were obtained by adding in quadrature the uncertainties in laboratory ($\delta\lambda_{\mathrm{lab}}$) and observed wavelength ($\delta\lambda_{\mathrm{obs}}$) calculated from the Gaussian/Lorentzian fit.
Some velocity uncertainties were anomalously high due to large uncertainties from the Kentucky database, and it is for this reason that the average photospheric and interstellar velocities were calculated by weighting each line velocity by its inverse error:
\begin{equation}\bar{v}=\frac{\sum_{i=1}^{N}\frac{v_{i}}{\delta{v_i}^2}}{\sum_{i=1}^{N}\frac{1}{\delta{v_i}^2}}\end{equation}
With associated error:
\begin{equation}\delta\bar{v}=\sqrt{\frac{1}{\sum_{i=1}^{N}\frac{1}{\delta{v_i}^2}}}\end{equation}
Where $v_i$ and $\delta{v_i}$ are the line velocities and their respective errors, and $N$ is the number of lines used to calculate the mean. 
The velocities were calculated using only lines with wavelength errors from Kentucky, and were observed in the STIS data. In our line survey, we identified three definitive velocity populations, one photospheric, and two interstellar. We measured the photospheric velocity as $23.8\pm{0.03}$km s$^{-1}$, while the interstellar velocities were measured as $19.4\pm{0.03}$km s$^{-1}$ and $8.64\pm{0.03}$km s$^{-1}$ respectively. In Table \ref{table:velocities}, we compare our photospheric velocity with that obtained by \citet{vennes01a}, and our interstellar velocities with those from \citet{redfield08a}. Hereafter, we refer to the interstellar velocities by the name of the cloud from which they appear to originate, as named by \citet{redfield08a}, where the $19.4\pm{0.03}$km s$^{-1}$ velocity corresponds to the Local Interstellar Cloud (LIC), and the $8.64\pm{0.03}$km s$^{-1}$ velocity to the Hyades Cloud. The velocities determined in this study appear to be in excellent agreement with those obtained by previous authors.

\begin{table}
\centering
\caption{A list of absorption features that could not be identified in our survey, where $\lambda_{\mathrm{Obs}}$ is the observed wavelength with accompanying error $\delta\lambda_{\mathrm{Obs}}$, and $W_{\lambda}$ is the equivalent width, with error $\delta{W_{\lambda}}$.}
\begin{tabular}[H]{@{}lccc}
\hline
$\lambda_{\mathrm{Obs}}$(\AA) & $\delta\lambda_{\mathrm{Obs}}$(m\AA) & $W_{\lambda}$(m\AA) & $\delta{W_{\lambda}}$(m\AA) \\
\hline
1171.277 & 1.9730 & 2.584 & 0.793 \\
1174.424 & 1.3261 & 5.417 & 0.843 \\
1186.174 & 2.8080 & 3.046 & 0.962 \\
1186.355 & 0.9279 & 9.953 & 0.927 \\
1196.733 & 1.7920 & 3.897 & 0.730 \\
1198.240 & 0.9257 & 6.856 & 0.636 \\
1199.443 & 1.0994 & 2.369 & 0.572 \\
1201.548 & 2.2597 & 1.689 & 0.543 \\
1204.561 & 2.0988 & 3.064 & 0.613 \\
1206.756 & 1.4513 & 3.354 & 0.686 \\
1206.812 & 2.4298 & 2.788 & 0.761 \\
1228.604 & 2.9613 & 3.232 & 0.663 \\
1232.311 & 3.4623 & 1.813 & 0.554 \\
1253.405 & 2.0101 & 1.727 & 0.472 \\
1255.177 & 3.0864 & 2.695 & 0.329 \\
1270.950 & 4.4021 & 4.710 & 0.558 \\
1274.017 & 1.8103 & 1.920 & 0.386 \\
1285.088 & 2.8810 & 1.148 & 0.359 \\
1291.912 & 2.3123 & 3.007 & 0.586 \\
1292.590 & 2.7298 & 2.880 & 0.578 \\
1295.987 & 2.6044 & 1.651 & 0.311 \\
1302.927 & 2.8485 & 3.624 & 0.653 \\
1318.082 & 1.0165 & 2.999 & 0.353 \\
1321.307 & 1.4181 & 2.564 & 0.403 \\
1322.416 & 1.3813 & 2.398 & 0.375 \\
1333.462 & 2.3978 & 2.265 & 0.613 \\
1442.574 & 2.4808 & 1.673 & 0.401 \\
1499.254 & 3.6779 & 3.035 & 0.827 \\
1513.608 & 3.2663 & 2.266 & 0.730 \\
\hline
\end{tabular}
\label{table:unidentified}
\end{table}

\begin{table}
\centering
\caption{A summary of the velocity populations identified in the spectrum. Each velocity was calculated using the weighted mean of the various line velocities measured in the STIS data. We also compare our determined photospheric velocity to that determined by \protect\citet{vennes01a}, and our ISM velocities with that of \protect\citet{redfield08a}.}
\begin{tabular}{@{}lcc}
\hline
Origin & $v_{\mathrm{Current}}$ (km s$^{-1}$) & $v_{\mathrm{Previous}}$ (km s$^{-1}$) \\
\hline
Photosphere & $23.8\pm{0.03}$ & $24.3\pm{1.7}$ \\
LIC & $19.4\pm{0.03}$ & $19.19\pm{0.09}$ \\
Hyades & $8.64\pm{0.03}$ & $8.61\pm{0.74}$ \\
\hline
\end{tabular}
\label{table:velocities}
\end{table}

\section{Model atmospheres and abundance determination}
 
To determine the abundances, we first used {\sc tlusty} version 200 to calculate a model atmosphere based on the abundances given by \citet{barstow03b} (C, N, O, Si, Fe, Ni) and \citet{vennes01a} (P, S) in NLTE. We fixed He/H to $1.00\times{10^{-5}}$, and $T_{\mathrm{eff}}$ and log $g$ to 52,500K and 7.53 respectively. The model atoms used by {\sc tlusty}, along with the number of levels included, are listed in Table \ref{table:modelatoms}. Next, as Al had not been included in a full NLTE model atmosphere calculation for G191-B2B before, we introduced the metal into the solution, and calculated a grid of models with varying Al abundances, the values of which are given in Table \ref{table:abungrid}. The model spectra were then synthesized using {\sc synspec} version 49, and Kurucz's 1992 line list as this has a complete set of oscillator strengths. It was interesting to note that even with the greatest abundance of Al, there was no noticable flux redistribution. We then determined the Al abundance using  {\sc xspec} \citep{arnaud96a}, which we will describe shortly. Upon determining the abundance, we recalculated the atmosphere with this value using {\sc tlusty}. 
As the abundances of C, N, O, Si, P, S, Fe and Ni are not so different from the values derived in \citet{barstow03b} and \citet{vennes01a}, any flux redistribution due to small abundance variations is likely to be a second order effect. Therefore, instead of calculating a model atmosphere for each metal, which is computationally expensive, we used {\sc synspec} 49 to modify the abundances, creating a model grid for each metal as given in Table \ref{table:abungrid}, again using the 1992 Kurucz line list. As with Al, we then used {\sc xspec} to determine the abundances.

{\sc xspec} calculates the abundances and associated errors by taking a model atmosphere grid and observational data as input, and interpolating between the different abundance values using a chi square ($\chi^{2}$) minimisation technique. {\sc xspec} has difficulty in performing computations with spectra containing many data points. With 60,000 data points, the STIS spectrum can not, therefore, be analysed in its entirety without dividing it into segments. Therefore, we extracted small regions of spectra containing the absorption features that we wish to analyse, summarised in Table \ref{table:linerange}. Isolating small sections of spectrum also has the advantage of minimising systematic errors due to poor normalisation of the continuum whilst fitting the abundances. We determined the abundance of each metal ionisation stage individually. In cases where there were multiple lines in an ionisation stage, we fixed the abundance to be the same for each feature. Photospheric features that appeared to be blended with additional components were modelled by including an additional Gaussian absorber using the {\sc xspec} model {\sc gabs} (see Appendix A for the exact parameterisation) as done by \citet{dickinson12c}. In all cases, we quote our formal errors to $1\sigma$ confidence, and assume one degree of freedom except where noted. We have tabulated our abundance determinations in Table \ref{table:obsabun}. We have also tabulated the parameter values and their respective errors obtained from fitting the {\sc gabs} component in Table \ref{table:circparam1}, along with the velocity of the absorber where relevant.

\begin{table}
\centering
\caption{Model atoms used in calculating the initial model atmosphere.}
\begin{tabular}{@{}llc}
\hline
Element & Ion & No of Levels \\
\hline
H & {\sc i} & 9 \\ 
He & {\sc i} & 24 \\
He & {\sc ii} & 20 \\ 
C & {\sc iii} & 23 \\
C & {\sc iv} & 41 \\ 
N & {\sc iii} & 32 \\
N & {\sc iv} & 23 \\ 
N & {\sc v} & 16 \\
O & {\sc iv} & 39 \\ 
O & {\sc v} & 40 \\
O & {\sc vi} & 20 \\ 
Si & {\sc iii} & 30 \\
Si & {\sc iv} & 23 \\ 
P & {\sc iv} & 14 \\
P & {\sc v} & 17 \\ 
S & {\sc iv} & 15 \\
S & {\sc v} & 12 \\ 
S & {\sc vi} & 16 \\
Fe & {\sc iv} & 43 \\ 
Fe & {\sc v} & 42 \\
Fe & {\sc vi} & 32 \\ 
Ni & {\sc iv} & 38 \\
Ni & {\sc v} & 48 \\ 
Ni & {\sc vi} & 42 \\
\hline
\end{tabular}
\label{table:modelatoms}
\end{table}

\begin{table*}
\centering
\caption{The grid of abundances used in fitting the metal absorption features. $T_{\mathrm{eff}}$ and log $g$ were fixed at 52500K and 7.53 respectively.}
\begin{tabular}[H]{@{}lllllllll}
\hline
C & N & O & Al & Si & P & S & Fe & Ni \\
\hline
$2.00\times{10}^{-8}$ & $2.00\times{10}^{-8}$ & $2.00\times{10}^{-8}$ & $2.00\times{10}^{-8}$ & $2.00\times{10}^{-8}$ & $2.00\times{10}^{-9}$ & $2.00\times{10}^{-8}$ & $2.00\times{10}^{-7}$ & $2.00\times{10}^{-8}$ \\
$6.00\times{10}^{-8}$ & $6.00\times{10}^{-8}$ & $6.00\times{10}^{-8}$ & $6.00\times{10}^{-8}$ & $6.00\times{10}^{-8}$ & $6.00\times{10}^{-9}$ & $6.00\times{10}^{-8}$ & $6.00\times{10}^{-7}$ & $6.00\times{10}^{-8}$ \\
$2.00\times{10}^{-7}$ & $2.00\times{10}^{-7}$ & $2.00\times{10}^{-7}$ & $2.00\times{10}^{-7}$ & $2.00\times{10}^{-7}$ & $2.00\times{10}^{-8}$ & $2.00\times{10}^{-7}$ & $2.00\times{10}^{-6}$ & $2.00\times{10}^{-7}$ \\
$6.00\times{10}^{-7}$ & $6.00\times{10}^{-7}$ & $6.00\times{10}^{-7}$ & $6.00\times{10}^{-7}$ & $6.00\times{10}^{-7}$ & $6.00\times{10}^{-8}$ & $6.00\times{10}^{-7}$ & $6.00\times{10}^{-6}$ & $6.00\times{10}^{-7}$ \\
$2.00\times{10}^{-6}$ & $2.00\times{10}^{-6}$ & $2.00\times{10}^{-6}$ & $2.00\times{10}^{-6}$ & $2.00\times{10}^{-6}$ & $2.00\times{10}^{-7}$ & $2.00\times{10}^{-6}$ & $2.00\times{10}^{-5}$ & $2.00\times{10}^{-6}$ \\
\hline
\end{tabular}
\label{table:abungrid}
\end{table*}

\begin{table*}
\centering
\caption{The list of lines examined to determine the abundances of the photospheric metals, and the spectral window extracted to perform the analysis.}
\begin{tabular}[H]{@{}lllclllc}
\hline
Ion & Wavelength (\AA) & $f$-value & Spectral Region (\AA) & Ion & Wavelength (\AA) & $f$-value & Spectral Region (\AA)\\
\hline
C  {\sc iii}  &  1174.9327  &  0.114   &  1174-1178  &  Fe {\sc iv}  &  1592.050  &  0.3341  &  1590-1605  \\
C  {\sc iii}  &  1175.263   &  0.274   &  1174-1178  &  Fe {\sc iv}  &  1601.652  &  0.3379  &  1590-1605  \\
C  {\sc iii}  &  1175.5903  &  0.069   &  1174-1178  &  Fe {\sc iv}  &  1603.177  &  0.2679  &  1590-1605  \\  
C  {\sc iii}  &  1175.7112  &  0.205   &  1174-1178  &  Fe {\sc v}   &  1280.470  &  0.0236  &  1280-1290  \\
C  {\sc iii}  &  1175.9871  &  0.091   &  1174-1178  &  Fe {\sc v}   &  1287.046  &  0.0363  &  1280-1290  \\
C  {\sc iii}  &  1176.3697  &  0.068   &  1174-1178  &  Fe {\sc v}   &  1288.172  &  0.0541  &  1280-1290  \\
C  {\sc iii}  &  1247.383   &  0.163   &  1245-1255  &  Fe {\sc v}   &  1293.382  &  0.0330  &  1290-1300  \\
C  {\sc iv}   &  1548.202   &  0.190   &  1545-1555  &  Fe {\sc v}   &  1297.549  &  0.0440  &  1290-1300  \\
C  {\sc iv}   &  1550.777   &  0.095   &  1545-1555  &  Fe {\sc v}   &  1311.828  &  0.1710  &  1305-1315  \\
N  {\sc iv}   &  1718.551   &  0.173   &  1715-1725  &  Fe {\sc v}   &  1320.409  &  0.1944  &  1320-1333  \\
N  {\sc v}    &  1238.821   &  0.156   &  1235-1245  &  Fe {\sc v}   &  1321.489  &  0.0905  &  1320-1333  \\
N  {\sc v}    &  1242.804   &  0.078   &  1235-1245  &  Fe {\sc v}   &  1323.271  &  0.1930  &  1320-1333  \\
O  {\sc iv}   &  1338.615   &  0.118   &  1334-1344  &  Fe {\sc v}   &  1330.405  &  0.2085  &  1320-1333  \\
O  {\sc iv}   &  1342.99    &  0.011   &  1334-1344  &  Fe {\sc v}   &  1331.189  &  0.0774  &  1320-1333  \\
O  {\sc iv}   &  1343.514   &  0.104   &  1334-1344  &  Fe {\sc v}   &  1331.639  &  0.1867  &  1320-1333  \\
Al {\sc iii}  &  1854.716   &  0.556   &  1850-1860  &  Ni {\sc iv}  &  1356.079  &  0.0963  &  1350-1360  \\
Al {\sc iii}  &  1862.79    &  0.277   &  1860-1870  &  Ni {\sc iv}  &  1398.193  &  0.3837  &  1395-1405  \\
Si {\sc iii}  &  1206.4995  &  1.610   &  1200-1210  &  Ni {\sc iv}  &  1399.947  &  0.2964  &  1395-1405  \\
Si {\sc iii}  &  1206.5551  &  1.640   &  1200-1210  &  Ni {\sc iv}  &  1400.682  &  0.2950  &  1395-1405  \\
Si {\sc iv}   &  1122.4849  &  0.819   &  1120-1130  &  Ni {\sc iv}  &  1411.451  &  0.3507  &  1410-1422  \\
Si {\sc iv}   &  1128.3248  &  0.0817  &  1120-1130  &  Ni {\sc iv}  &  1416.531  &  0.1949  &  1410-1422  \\
Si {\sc iv}   &  1128.3400  &  0.736   &  1120-1130  &  Ni {\sc iv}  &  1419.577  &  0.1309  &  1410-1422  \\
Si {\sc iv}   &  1393.7546  &  0.508   &  1390-1400  &  Ni {\sc iv}  &  1421.216  &  0.2863  &  1410-1422  \\
Si {\sc iv}   &  1402.7697  &  0.252   &  1400-1410  &  Ni {\sc iv}  &  1430.190  &  0.1794  &  1425-1435  \\
P  {\sc iv}   &  950.657    &  1.470   &  945-955    &  Ni {\sc iv}  &  1432.449  &  0.1253  &  1425-1435  \\
P  {\sc v}    &  1117.977   &  0.467   &  1115-1125  &  Ni {\sc iv}  &  1452.220  &  0.3596  &  1450-1460  \\
P  {\sc v}    &  1128.008   &  0.231   &  1125-1135  &  Ni {\sc iv}  &  1498.893  &  0.1618  &  1495-1505  \\
S  {\sc iv}   &  1062.662   &  0.052   &  1060-1070  &  Ni {\sc v}   &  1230.435  &  0.2649  &  1230-1240  \\
S  {\sc iv}   &  1072.974   &  0.045   &  1070-1080  &  Ni {\sc v}   &  1232.807  &  0.1764  &  1230-1240  \\
S  {\sc vi}   &  933.378    &  0.433   &  930-940    &  Ni {\sc v}   &  1233.257  &  0.1605  &  1230-1240  \\
S  {\sc vi}   &  944.523    &  0.213   &  940-950    &  Ni {\sc v}   &  1234.393  &  0.1334  &  1230-1240  \\
Fe {\sc iv}   &  1542.155   &  0.1386  &  1540-1550  &  Ni {\sc v}   &  1235.831  &  0.1982  &  1230-1240  \\
Fe {\sc iv}   &  1542.697   &  0.2818  &  1540-1550  &  Ni {\sc v}   &  1236.277  &  0.1094  &  1230-1240  \\
Fe {\sc iv}   &  1544.486   &  0.2511  &  1540-1550  &  Ni {\sc v}   &  1239.552  &  0.1116  &  1230-1240  \\
Fe {\sc iv}   &  1546.404   &  0.2070  &  1540-1550  &  Ni {\sc v}   &  1241.627  &  0.2003  &  1240-1250  \\
Fe {\sc iv}   &  1562.751   &  0.2032  &  1560-1572  &  Ni {\sc v}   &  1243.504  &  0.0815  &  1240-1250  \\
Fe {\sc iv}   &  1568.276   &  0.3012  &  1560-1572  &  Ni {\sc v}   &  1243.662  &  0.1194  &  1240-1250  \\
Fe {\sc iv}   &  1569.222   &  0.1291  &  1560-1572  &  Ni {\sc v}   &  1244.027  &  0.0503  &  1240-1250  \\
Fe {\sc iv}   &  1570.178   &  0.3147  &  1560-1572  &  Ni {\sc v}   &  1245.176  &  0.2348  &  1240-1250  \\
Fe {\sc iv}   &  1570.416   &  0.2741  &  1560-1572  &               &            &          &             \\
\hline
\end{tabular}
\label{table:linerange}
\end{table*}

\begin{table}
\centering
\caption{Summary of the abundances determined in this work, along with $1\sigma$ errors.}
\begin{tabular}[H]{@{}llll}
\hline
Ion & Abundance & $-1\sigma$ & $+1\sigma$ \\
\hline
C {\sc iii}  & $1.72\times{10}^{-7}$ & $0.02\times{10}^{-7}$ & $0.02\times{10}^{-7}$ \\
C {\sc iv}   & $2.13\times{10}^{-7}$ & $0.15\times{10}^{-7}$ & $0.29\times{10}^{-7}$ \\
N {\sc iv}   & $1.58\times{10}^{-7}$ & $0.14\times{10}^{-7}$ & $0.14\times{10}^{-7}$ \\
N {\sc v}    & $2.16\times{10}^{-7}$ & $0.04\times{10}^{-7}$ & $0.09\times{10}^{-7}$ \\
O {\sc iv}   & $4.12\times{10}^{-7}$ & $0.08\times{10}^{-7}$ & $0.08\times{10}^{-7}$ \\
Al {\sc iii} & $1.60\times{10}^{-7}$ & $0.08\times{10}^{-7}$ & $0.07\times{10}^{-7}$ \\
Si {\sc iii} & $3.16\times{10}^{-7}$ & $0.30\times{10}^{-7}$ & $0.31\times{10}^{-7}$ \\
Si {\sc iv}  & $3.68\times{10}^{-7}$ & $0.14\times{10}^{-7}$ & $0.13\times{10}^{-7}$ \\
P {\sc iv}   & $8.40\times{10}^{-8}$ & $1.18\times{10}^{-8}$ & $1.18\times{10}^{-8}$ \\
P {\sc v}    & $1.64\times{10}^{-8}$ & $0.02\times{10}^{-8}$ & $0.02\times{10}^{-8}$ \\
S {\sc iv}   & $1.71\times{10}^{-7}$ & $0.02\times{10}^{-7}$ & $0.02\times{10}^{-7}$ \\
S {\sc vi}   & $5.23\times{10}^{-8}$ & $0.13\times{10}^{-8}$ & $0.10\times{10}^{-8}$ \\
Fe {\sc iv}  & $1.83\times{10}^{-6}$ & $0.03\times{10}^{-6}$ & $0.03\times{10}^{-6}$ \\
Fe {\sc v}   & $5.00\times{10}^{-6}$ & $0.06\times{10}^{-6}$ & $0.06\times{10}^{-6}$ \\
Ni {\sc iv}  & $3.24\times{10}^{-7}$ & $0.05\times{10}^{-7}$ & $0.13\times{10}^{-7}$ \\
Ni {\sc v}   & $1.01\times{10}^{-6}$ & $0.03\times{10}^{-6}$ & $0.03\times{10}^{-6}$ \\
\hline
\end{tabular}
\label{table:obsabun}
\end{table}

\begin{table*}
\centering
\caption{The best fitting {\sc xspec} Gaussian model parameters for a given transition, where $\lambda_{\mathrm{lab}}$ is the lab wavelength, $E_l$ is the centroid wavelength of the Gaussian in keV, $\sigma_l$ is the line width of the Gaussian in keV, $\tau_l$ is the strength, and $v$ is the corresponding velocity of the absorber. All parameters have been calculated to $1\sigma$ confidence.}
\begin{tabular}[H]{@{}lllll}
\hline
Ion          & Parameter & Value & $-1\sigma$ & $+1\sigma$ \\
\hline
C {\sc iv}   & $\lambda_{\mathrm{lab}}$ (\AA) & 1548.203 & - & - \\
             & $E_l$ ($10^{-3}$keV) & 8.008122 & 0.0000047 & 0.0000038 \\
             & $\sigma_l$ ($10^{-7}$keV) & 1.34141 & 0.03611 & 0.03689 \\
             & Strength ($10^{-7}$) & 1.19420 & 0.0426 & 0.0489 \\
             & $\lambda_{\mathrm{obs}}$ (\AA) & 1548.246 & 0.000735 & 0.000909 \\
             & $v$ (km s$^{-1}$) & 8.26 & 0.14 & 0.18 \\
\hline
C {\sc iv}   & $\lambda_{\mathrm{lab}}$ (\AA) & 1550.777 & - & - \\
             & $E_l$ ($10^{-3}$keV) & 7.994829 & 0.0000036 & 0.0000040 \\
             & $\sigma_l$ ($10^{-7}$keV) & 1.26307 & 0.03597 & 0.02623 \\
             & Strength ($10^{-7}$) & 7.18607 & 0.20447 & 0.17263 \\
             & $\lambda_{\mathrm{obs}}$ (\AA) & 1550.820 & 0.000776 & 0.000698 \\
             & $v$ (km s$^{-1}$) & 8.30 & 0.15 & 0.13 \\
\hline
Si {\sc iii} & $\lambda_{\mathrm{lab}}$ (\AA) & 1206.4995 & - & - \\
             & $E_l$ ($10^{-3}$keV) & 10.276140 & 0.0000070 & 0.0000020 \\
             & $\sigma_l$ ($10^{-7}$keV) & 0.925806 & 0.026966 & 0.025814 \\
             & Strength ($10^{-7}$) & 4.79551 & 0.13501 & 0.14099 \\
             & $\lambda_{\mathrm{obs}}$ (\AA) & 1206.537 & 0.000235 & 0.000822 \\
             & $v$ (km s$^{-1}$) & 9.24 & 0.06 & 0.20 \\
\hline
Si {\sc iv}  & $\lambda_{\mathrm{lab}}$ (\AA) & 1393.7546 & - & - \\
             & $E_l$ ($10^{-3}$keV) & 8.895525 & 0.0000277 & 0.0000208 \\
             & $\sigma_l$ ($10^{-7}$keV) & 0.926033 & 0.183263 & 0.293867 \\
             & Strength ($10^{-7}$) & 0.434293 & 0.072113 & 0.098237 \\
             & $\lambda_{\mathrm{obs}}$ (\AA) & 1393.795 & 0.003259 & 0.004340 \\
             & $v$ (km s$^{-1}$) & 8.73 & 0.70 & 0.93 \\
\hline
Si {\sc iv}  & $\lambda_{\mathrm{lab}}$ (\AA) & 1402.7697 & - & - \\
             & $E_l$ ($10^{-3}$keV) & 8.838369 & 0.0000503 & 0.0000411 \\
             & $\sigma_l$ ($10^{-7}$keV) & 0.890117 & 0.341027 & 0.550783 \\
             & Strength ($10^{-7}$) & 0.206168 & 0.069738 & 0.087402 \\
             & $\lambda_{\mathrm{obs}}$ (\AA) & 1402.809 & 0.006523 & 0.007984 \\
             & $v$ (km s$^{-1}$) & 8.31 & 1.39 & 1.71 \\
\hline
\end{tabular}
\label{table:circparam1}
\end{table*}

\subsection{Carbon}
\begin{figure*}
\begin{centering}
\includegraphics[width=100mm]{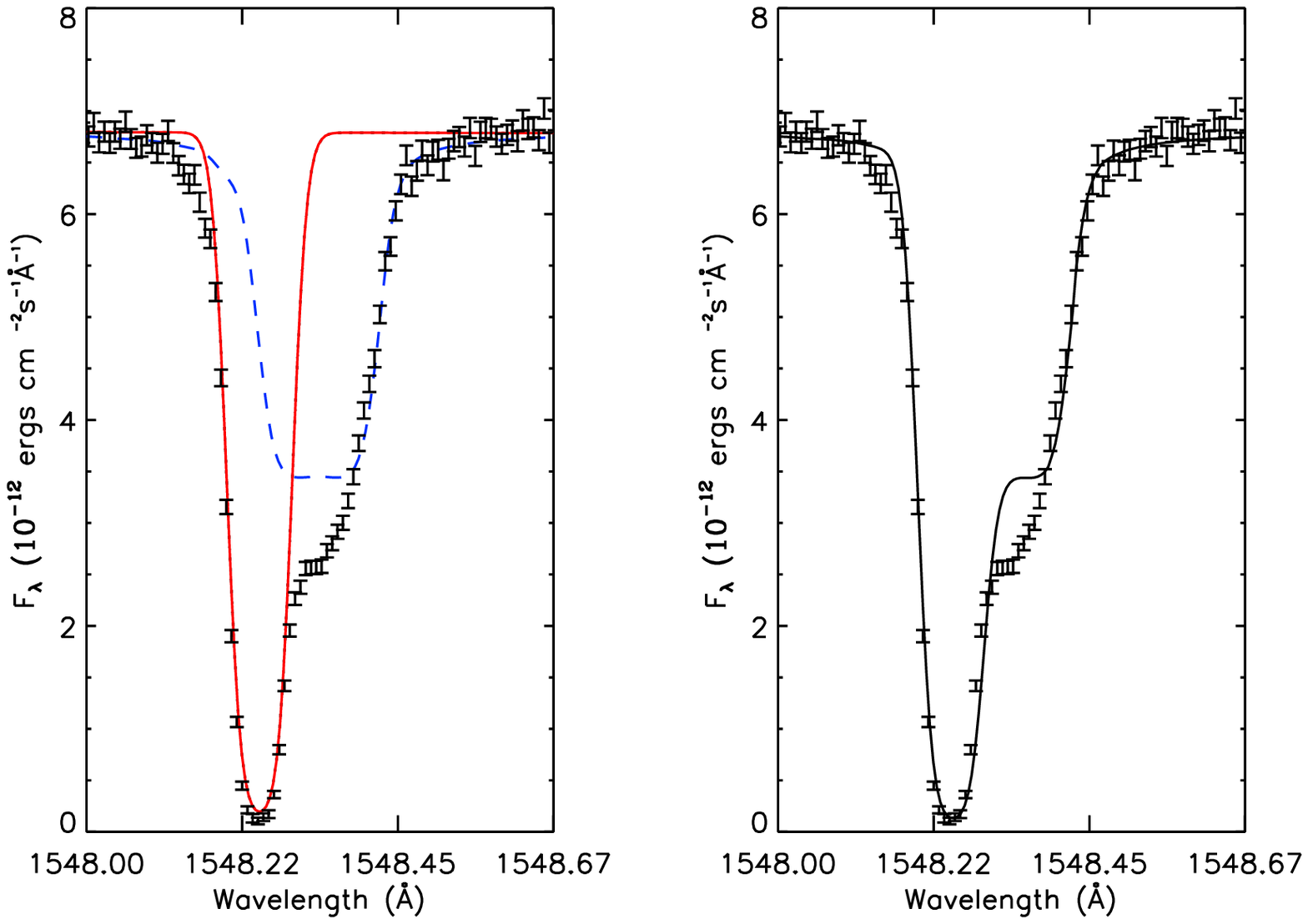}
\caption{Addition of a Gaussian component (red solid line) to the photospheric (blue dashed line) component of C {\sc iv} 1548\AA. The left subplot shows the individual contributions to the absorption feature, while the right subplot shows the combined contribution from both lines. C {\sc iv}/H=$2.13\times{10^{-7}}$. The discrepancy is explained in text.}
\label{fig:civ1548circ}
\end{centering}
\end{figure*}

\begin{figure*}
\begin{centering}
\includegraphics[width=100mm]{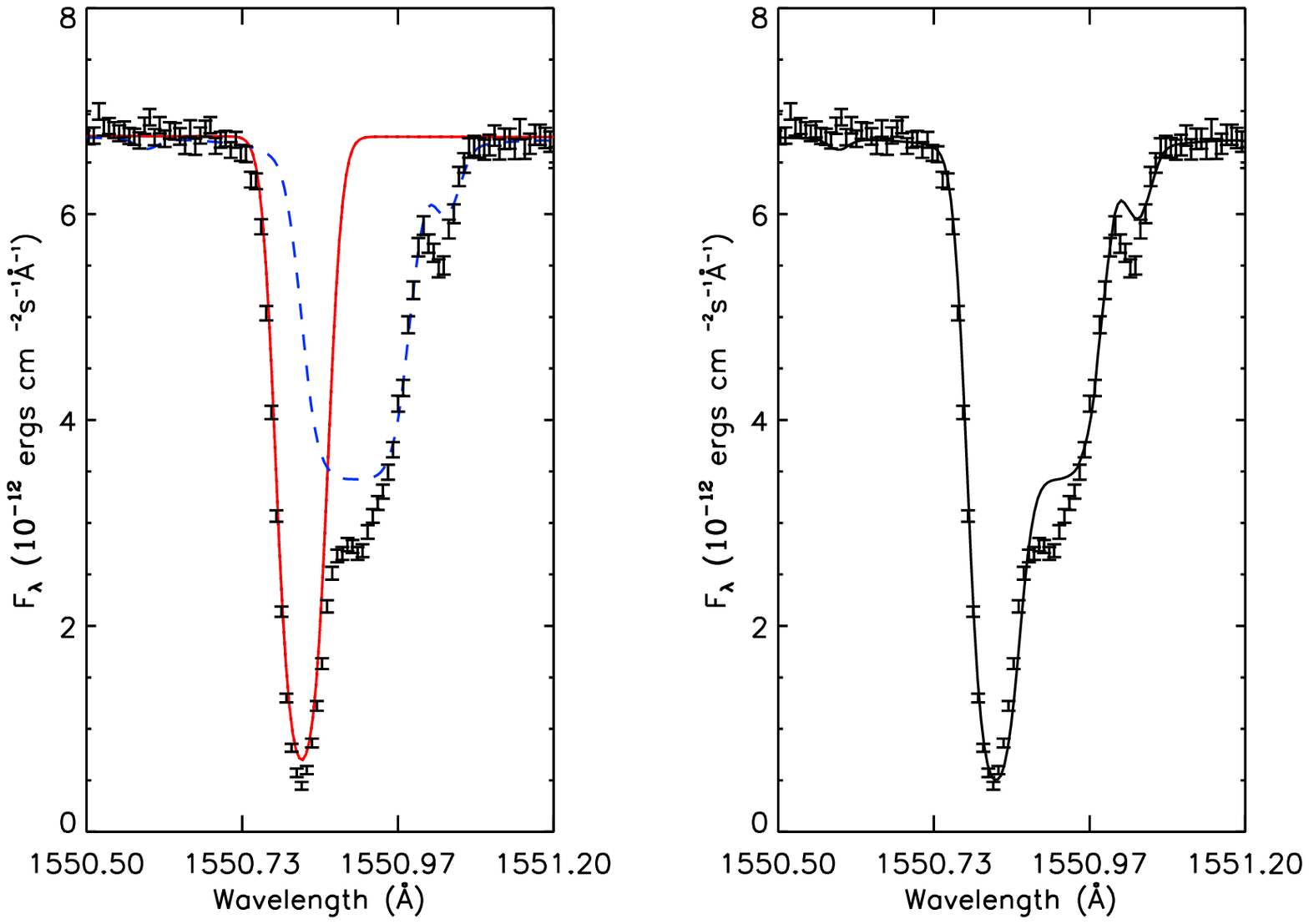}
\caption{The same as Figure \ref{fig:civ1548circ}, but for the C {\sc iv} 1550\AA\, line.}
\label{fig:civ1550circ}
\end{centering}
\end{figure*}

As discussed in the introduction, C {\sc iv} 1548 and 1550\AA\, in G191-B2B's photospheric spectrum are well documented, each being blended with a circumstellar absorption feature. Including a Gaussian, we obtain C {\sc iv}/H=$2.13_{-0.15}^{+0.29}\times{10^{-7}}$ (cf. Figures \ref{fig:civ1548circ} and \ref{fig:civ1550circ}. We can also see in Figures \ref{fig:civ1548circ} and \ref{fig:civ1550circ} that there is a small discrepancy between the predicted and observed absorption profiles. This "shelf" is not observed in \citet{vennes01a}'s analysis of the C {\sc iv} profiles, however, this is likely due to the lower resolution of the data, which was obtained using the E140M grating. In both cases, we believe the shelf arises due to the presence of Ni {\sc iv} transitions at 1548.220 and 1550.777\AA\, that have poor oscillator strength determinations.

\begin{figure}
\begin{centering}
\includegraphics[width=80mm]{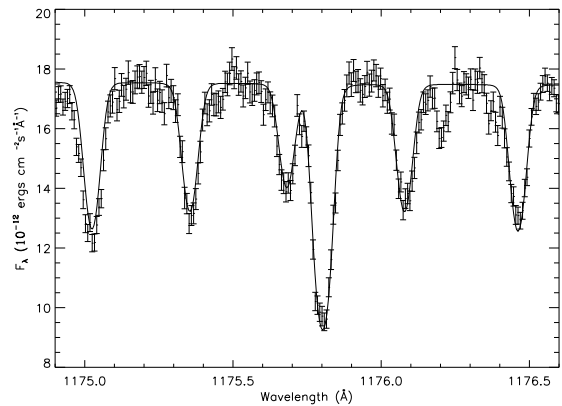}
\caption{Comparison between the model and observed spectra for the C {\sc iii} sextuplet spanning 1174-1177\AA\, with C {\sc iii}/H=$1.72\times{10^{-7}}$.}
\label{fig:ciii1174}
\end{centering}
\end{figure}

As well as the C {\sc iii} lines listed in Table \ref{table:linerange}, the C {\sc iii} resonant transition at 977.0201\AA\, (hereafter C {\sc iii} 977\AA) was available to fit, however, attempts to do so were unsuccessful, as the predicted profile could not descend deeply enough relative to the continuum, and increasing the C abundance resulted in large pressure broadening that was uncharacteristic of the observed profile. This does not come as a surprise, as the C {\sc iii} 977\AA\, line has been observed along the line of sight to G191-B2B by \citet{lehner03a} in the ISM. We attempted to add a Gaussian to account for this, but again, a satisfactory fit could not be obtained. Therefore, we determined the C {\sc iii} abundance using only the lines in Table \ref{table:linerange}, obtaining C {\sc iii}/H=$1.72_{-0.02}^{+0.02}\times{10^{-7}}$ (cf. Figure \ref{fig:ciii1174} for the C {\sc iii} sextuplet).

Our C {\sc iii} and C {\sc iv} abundances are in good agreement with \citet{barstow03b}'s C {\sc iii} value of $1.99_{-0.88}^{+0.44}\times{10}^{-7}$, but not with their C {\sc iv} value of $4.00_{-0.98}^{+0.44}\times{10}^{-7}$. We note, however, that \citet{barstow03b}'s C {\sc iv} abundance was obtained without taking the circumstellar absorption into account. We also calculated the velocity of the circumstellar lines, obtaining $8.26_{-0.14}^{+0.18}$ and $8.30_{-0.15}^{+0.13}$ km s$^{-1}$ for the C {\sc iv} 1548 and 1550\AA\, lines respectively, both in relatively good agreement with our obtained velocity corresponding to the Hyades Cloud velocity as identified by \citet{redfield08a}. 

\subsection{Nitrogen}
Many N {\sc iv} transitions exist in the \textit{FUSE} spectrum of G191-B2B from 915-930\AA, but we neglected using these lines in favour of N {\sc iv} 1718.551\AA\, (hereafter N {\sc iv} 1718\AA) due to the higher S/N. We obtained N {\sc iv}/H=$1.58_{-0.14}^{+0.14}\times{10^{-7}}$.

Using the resonant N {\sc v} doublet lines at 1238.821 and 1242.804\AA, we obtain N {\sc v}/H=$2.16_{-0.04}^{+0.09}\times{10^{-7}}$. Our N {\sc iv} abundance is in good agreement with that obtained by \citet{barstow03b} of $1.60_{-0.21}^{+0.41}\times{10^{-7}}$.

\subsection{Oxygen}
Absorption features of O {\sc iv}-{\sc vi} were detected in our line survey. Given the temperature of G191-B2B, the equivalent widths of the resonant O {\sc vi} transitions (1031.93 and 1037.62\AA) are quite weak, and were neglected from our analysis. Futhermore, attempts to obtain an abundance with the well known O {\sc v} 1371.296\AA\, line were not possible. The predicted profile of the O {\sc v} line did not appear in our synthesised spectrum unless O abundances of $\approx{10^{-6}}$ were specified. This issue may be related to that noted by \citet{vennes00a}, whereby the ionisation fraction for oxygen is poorly calculated for O {\sc iv}/O {\sc v}. \citet{vennes00a} ruled out $T_{\mathrm{eff}}$ and log $g$ variations as being the cause, suggesting reasons such as additional atmospheric consituents, stratification, or inadequate model atom treatment. This is addressed in \citet{vennes01a}, whereby they used the O abundance determined from the O {\sc iv} lines. This may also explain the large uncertainty in the O abundance value calculated by \citet{barstow03b} ($3.51_{-2.00}^{+7.40}\times{10^{-7}}$). Our abundance determination therefore is based only upon the O {\sc iv} triplet listed in Table \ref{table:linerange}, whereby we found O {\sc iv}/H=$4.12_{-0.08}^{+0.08}\times{10^{-7}}$, in good agreement with \citet{barstow03b}.

\subsection{Aluminium}
\begin{figure}
\begin{centering}
\includegraphics[width=80mm]{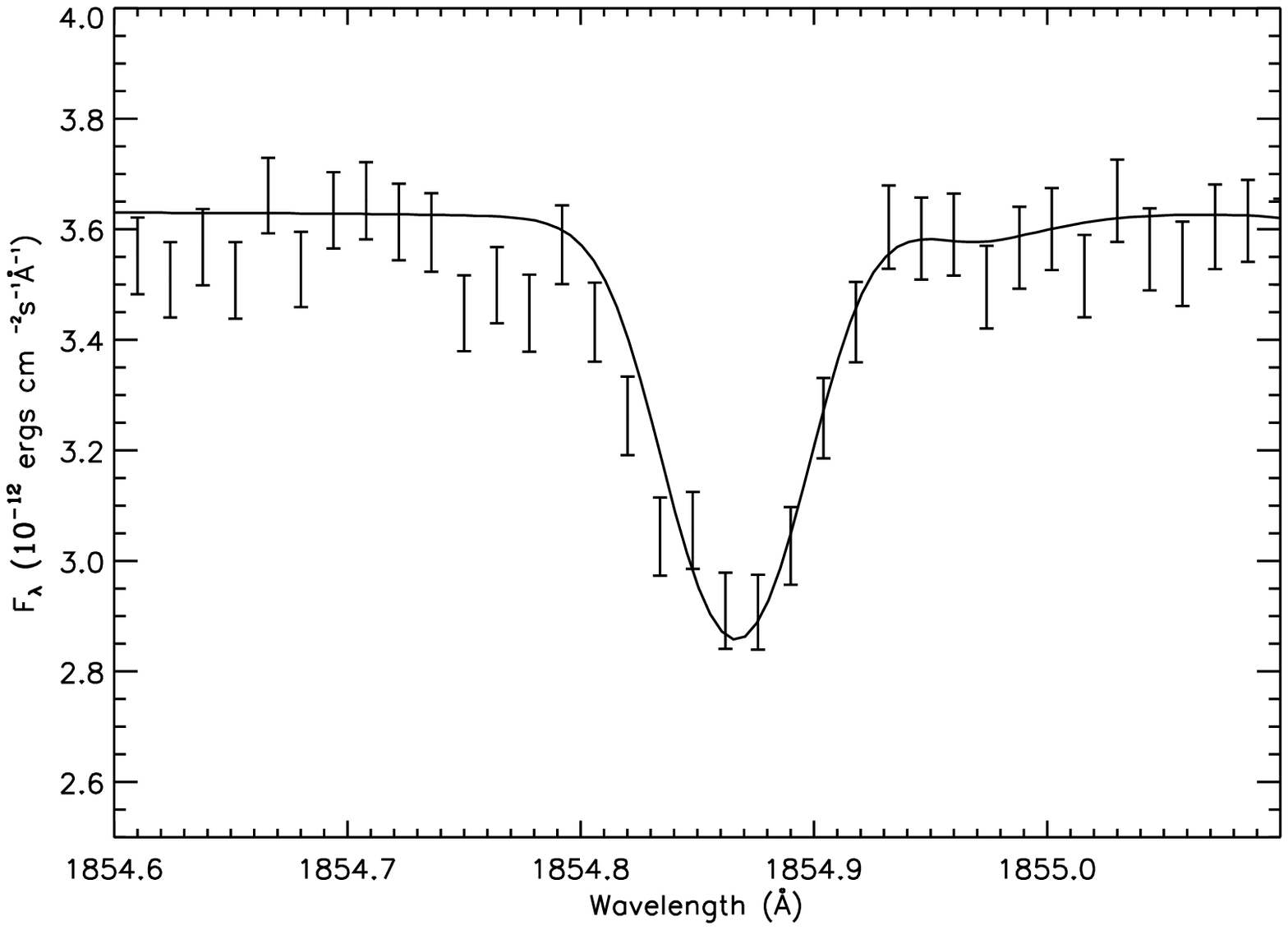}
\includegraphics[width=80mm]{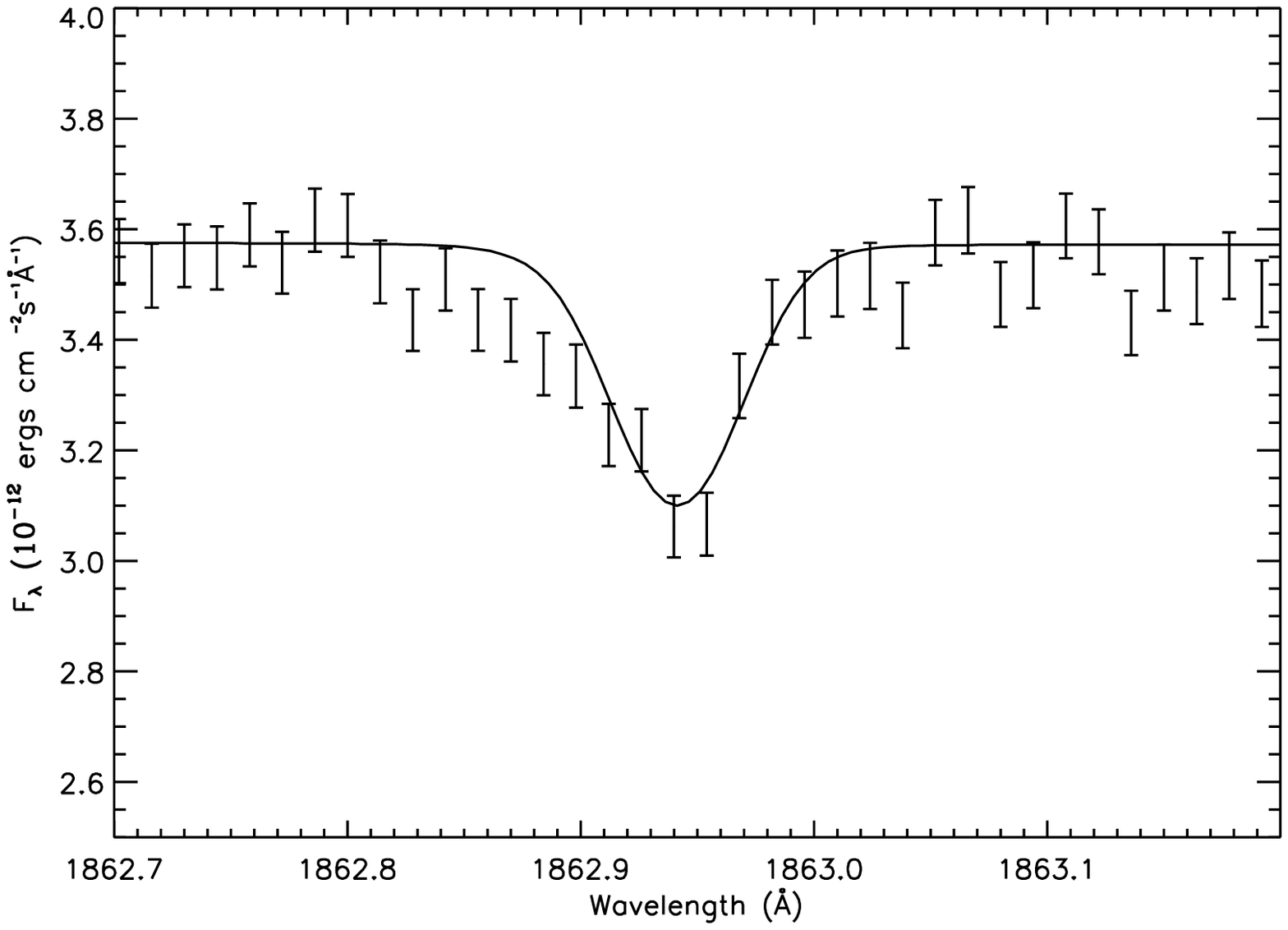}
\caption{Comparison between the model (solid line) and observed (error bars) spectra for the Al {\sc iii} 1854.716 (top plot) and 1862.79\AA\, (bottom plot) lines, with Al {\sc iii}/H=$1.60\times{10^{-7}}$.}
\label{fig:alfigs}
\end{centering}
\end{figure}

Al was first observed in G191-B2B by \citet{holberg98b} using \textit{IUE}, however, an abundance was not calculated. \citet{holberg03a}, then considered the abundance of Al assuming LTE, arriving at an estimated value of $3.02\times{10}^{-7}$. We do not observe other isolated photospheric Al features. We obtained Al {\sc iii}/H=$1.60_{-0.08}^{+0.07}\times{10^{-7}}$ (cf. Figure \ref{fig:alfigs}).

\subsection{Silicon}
\begin{figure*}
\begin{centering}
\includegraphics[width=100mm]{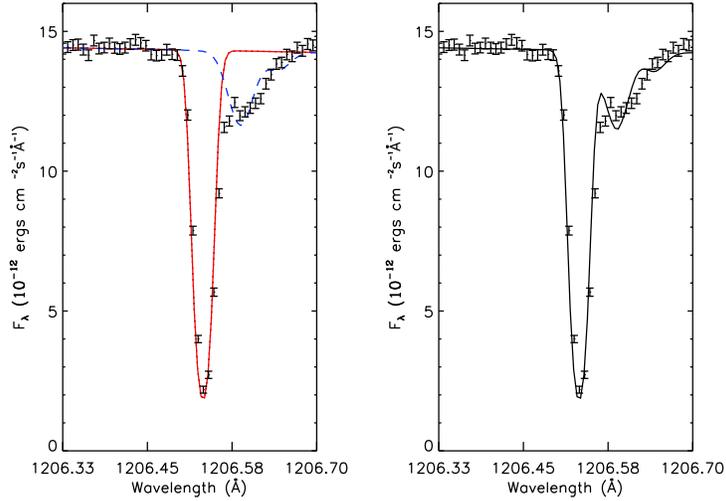}
\caption{Addition of a Gaussian component (red solid line) to the photospheric (blue dashed line) Si {\sc iii} 1206\AA. The left subplot shows the individual contributions to the absorption feature, while the right subplot shows the combined contribution from both lines. Si {\sc iii}/H=$3.16\times{10^{-7}}$.}
\label{fig:siiii1206circ}
\end{centering}
\end{figure*}

For Si {\sc iii} 1206.4995\AA\, (hereafter Si {\sc iii} 1206\AA), we were unable to fit the absorption feature as the predicted line profile did not descend deeply enough relative to the continuum. Increases in Si abundance produced a line profile that was pressure broadened far beyond that observed. Including a Gaussian into the fit, we obtained Si {\sc iii}/H=$3.16_{-0.30}^{+0.31}\times{10^{-7}}$ (see Figure \ref{fig:siiii1206circ}). We also determined the velocity of the Gaussian to be $9.24_{-0.06}^{+0.20}$ km s$^{-1}$. While we note that this velocity is not encompassed by the Hyades cloud velocity, the two values are quite close.

\begin{figure*}
\begin{centering}
\includegraphics[width=100mm]{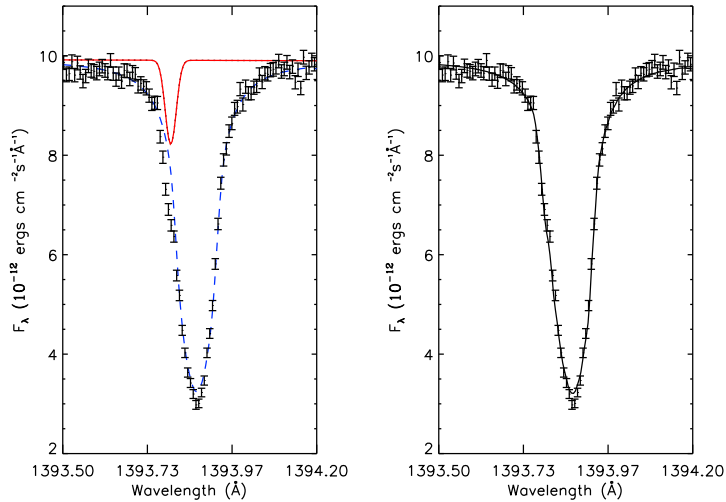}
\caption{Same as Figure \ref{fig:siiii1206circ}, but for the Si {\sc iv} 1393\AA\, line with Si {\sc iv}/H=$3.68\times{10^{-7}}$.}
\label{fig:siiv1393circ}
\end{centering}
\end{figure*}

\begin{figure*}
\begin{centering}
\includegraphics[width=100mm]{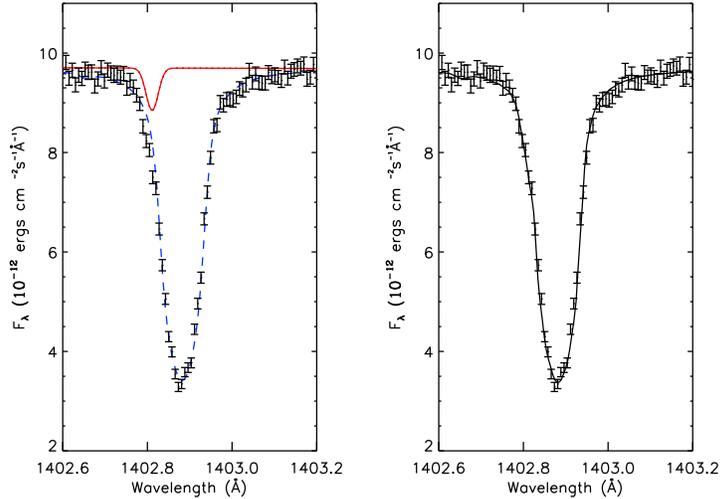}
\caption{Same as Figure \ref{fig:siiv1393circ}, but for the Si {\sc iv} 1402\AA\, line.}
\label{fig:siiv1402circ}
\end{centering}
\end{figure*}

\citet{holberg03a} noted an asymmetry in the absorption profiles of Si {\sc iv} 1393.7546 and 1402.7697\AA\, (hereafter Si {\sc iv} 1393\AA\, and 1402\AA\, respectively), observing a blue shifted component relative to the photospheric velocity. We confirmed this observation and included a Gaussian to account for its presence. We also included the excited Si IV transitions 1122.4849, 1128.3248, and 1128.3400\AA\, in order to further constrain the abundance. We derived Si {\sc iv}/H=$3.68_{-0.14}^{+0.13}\times{10^{-7}}$. The velocities of the two Gaussians (see Figures \ref{fig:siiv1393circ} and \ref{fig:siiv1402circ}) are $8.73_{-0.70}^{+0.93}$ and $8.31_{-1.39}^{+1.71}$ km s$^{-1}$ for the Si {\sc iv} 1393 and 1402\AA lines respectively, in good agreement with the Hyades cloud velocity. Both of our determined Si abundances are not encompassed by \citet{barstow03b}'s value of $8.65_{-3.50}^{+3.20}\times{10^{-7}}$, but the difference is not very large.

\subsection{Phosphorus}
Resonance absorption features of P have been observed in a handful of DAs, including GD\,394 \citep{chayer00a}, GD\,71, REJ\,1918+595 and REJ\,0605-482 \citep{dobbie05a}. P was first observed in G191-B2B by \citet{vennes96a} using the \textit{ORFEUS} telescope, deriving an LTE abundance of $2.51_{-0.93}^{+1.47}\times{10^{-8}}$. Determination of the P {\sc iv} abundance was complicated due to the proximity of the line to the Lyman $\gamma$ centroid, where the continuum varies rapidly with wavelength. We determined P {\sc iv}/H=$8.40_{-1.18}^{+1.18}\times{10^{-8}}$ and P {\sc v}/H=$1.64_{-0.02}^{+0.02}\times{10^{-8}}$. Our P {\sc v} abundance appears to be in closer agreement with that obtained by \citet{vennes96a}.

\subsection{Sulphur}
This metal was also detected in G191-B2B by \citet{vennes96a} with the \textit{ORPHEUS} telescope, deriving an LTE abundance of S/H=$3.16_{-1.58}^{+3.15}\times{10^{-7}}$. We calculated S {\sc iv}/H=$1.71_{-0.02}^{+0.02}\times{10^{-7}}$ and S {\sc vi}=$5.23_{-0.13}^{+0.10}\times{10^{-8}}$. Our S {\sc iv} abundance is also in agreement with \citet{vennes96a}.

\subsection{Iron}
\begin{figure}
\begin{centering}
\includegraphics[width=80mm]{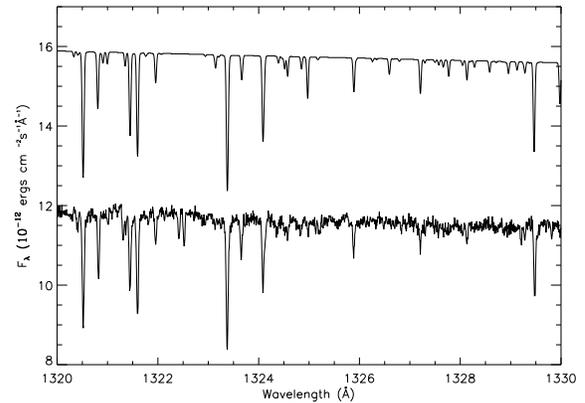}
\caption{Comparison between the model (top line) and observed (bottom line) spectra for several Fe {\sc v} lines, with Fe {\sc v}/H=$5.00\times{10^{-6}}$. The synthetic spectrum is offset for clarity.}
\label{fig:fe1330}
\end{centering}
\end{figure}

While there have been many measurements of the Fe content in various white dwarf photospheric spectra \citep{barstow03b,vennes06b}, none have used the same selection of lines, due to the large choice available. Here, we chose to use 12 strong lines for each ion listed in Table \ref{table:linerange}. We calculated Fe {\sc iv}/H=$1.83_{-0.03}^{+0.03}\times{10^{-6}}$ and Fe {\sc v}/H=$5.00_{-0.06}^{+0.06}\times{10^{-6}}$ (cf. Figure \ref{fig:fe1330}), with our Fe {\sc v} abundance in agreement with \citet{barstow03b}'s value of $3.30_{-1.20}^{+3.10}\times{10^{-6}}$.

\subsection{Nickel}
\begin{figure}
\begin{centering}
\includegraphics[width=80mm]{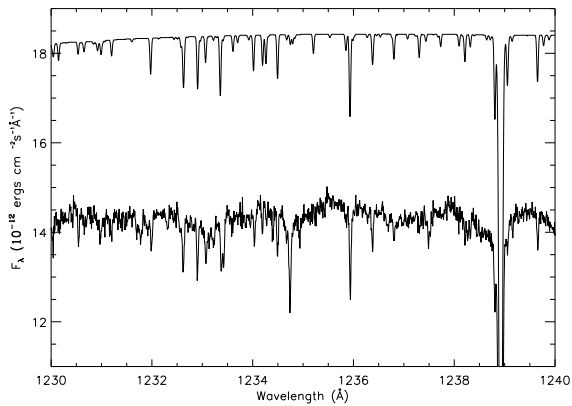}
\caption{Comparison between the model (top line) and observed (bottom line) spectra for several Ni {\sc v} lines, with Ni {\sc v}/H=$1.01\times{10^{-6}}$. The synthetic spectrum is offset for clarity. The large absorption feature near 1239\AA\, is from N {\sc v}.}
\label{fig:ni1230}
\end{centering}
\end{figure}

Like Fe, Ni is a complicated atomic system with many transitions. We use a similar method to Fe to determine the abundance, obtaining Ni {\sc iv}/H=$3.24_{-0.05}^{+0.13}\times{10^{-7}}$ and Ni {\sc v}/H=$1.01_{-0.03}^{+0.03}\times{10^{-6}}$ (cf. Figure \ref{fig:ni1230}, where our Ni {\sc iv} abundance agrees with \citet{barstow03b}'s value of $2.40_{-0.24}^{+0.84}\times{10^{-7}}$.

\section{Discussion}
Our line survey has been successful in providing new line identifications, with over 95 percent of the absorption features detected in the STIS spectra and 100 percent of those in the \textit{FUSE} data accounted for. Curiously, we confirmed the detection made by \citet{vennes05a} of Ge {\sc iv}, but we made no detections of heavier metals, or metals with atomic number between S and Fe. 

In particular, we made no detections of Cr, Mn, or Co, for which \citet{holberg03a} gave an abundance limit of $10^{-8}$.
However, as highlighted in the previous section, some discrepancies still remain regarding predicted line profiles and ionisation balance for particular metals. In our discussion of the results, we first consider the metals where we have potentially identified circumstellar absorption features. Next, we discuss the calculated abundances of G191-B2B, and compare them to the solar abundances of \citet{asplund09a}, and the predicted abundances of \citet{chayer94a,chayer95a,chayer95b} from radiative levitation calculations (cf. Figure \ref{fig:abunplot}). Finally, we consider the new atomic data, and the potential impact this may have on future model atmosphere calculations through the additional opacity that needs to be included.

\subsection{Circumstellar absorption}
As stated in section 4.1, the C {\sc iv} 1548 and 1550\AA\, lines are accompanied by circumstellar absorption features which have been thoroughly documented. We have also confirmed the detection of circumstellar absorption in Si {\sc iv}. The method of including a Gaussian to account for the circumstellar component is not new. \citet{dickinson12c} accounted for the line shape of the photospheric profile (which may include re-emission) as well as the circumstellar component, giving a more accurate representation of the absorption. However, our analysis of the profile allowed us to derive velocities and to make a comparison between the Hyades cloud and the circumstellar velocities. 

\citet{bannister03a} noted the similarity between the ISM and circumstellar velocities along the line of sight to G191-B2B, and suggested that the circumstellar lines arose from material ionised by the str\"{o}mgren sphere of the white dwarf. \citet{redfield08a} made the association between this ISM component with that of the Hyades cloud. This hypothesis was supported by \citet{dickinson12b}'s findings, and suggested that the Hyades cloud fell inside G191-B2B's str\"{o}mgren sphere. Our velocity measurements of the C {\sc iv} and Si {\sc iv} circumstellar features encompass the Hyades cloud velocity, and appear to support the str\"{o}mgren sphere hypothesis. The velocity and uncertainty of the Si {\sc iii} 1206\AA\, line does not quite encompass the Hyades cloud velocity. However, the discrepancy is small. So we can reasonably associate the Si {\sc iii} line with the Hyades cloud. The issue of the C {\sc iii} 977\AA\, line is more difficult owing to the low resolution of the data. While we can infer that there is additional absorption, we cannot derive any information about the absorption profile, and hence cannot derive a reliable velocity. The inability to fit a profile is most likely due to the resolution of the \textit{FUSE} spectrometer.

\subsection{General abundance discussion}
\begin{figure*}
\begin{centering}
\includegraphics[width=120mm]{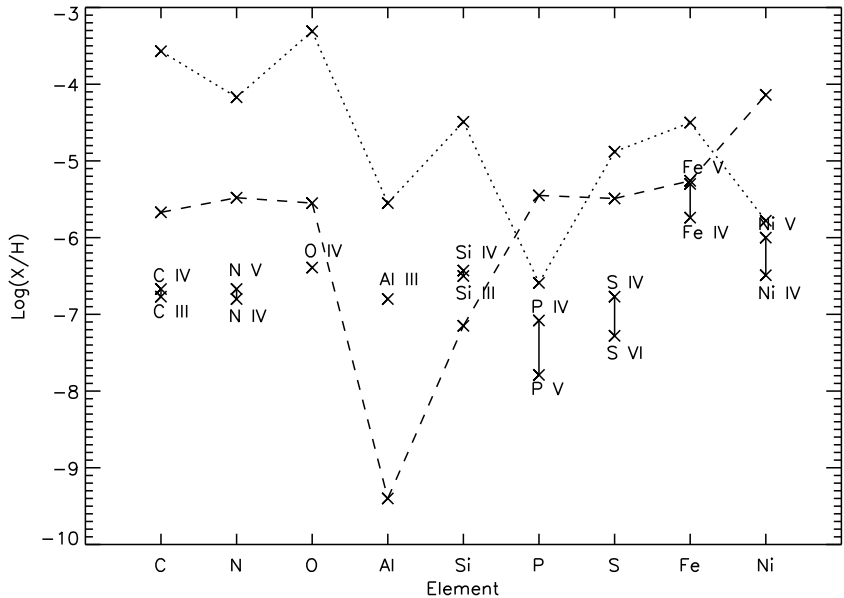}
\caption{A comparison of abundances with respect to H for solar abundance \citep{asplund09a} (dotted line), atmospheric abundance as predicted from radiative levitation (dashed line) \citep{chayer94a,chayer95a,chayer95b}, and the atmospheric abundances determined in our analysis (solid line). The error bars are omitted for clarity.}
\label{fig:abunplot}
\end{centering}
\end{figure*}

With the exceptionally high S/N of the \textit{FUSE} and STIS spectra, we have been able to constrain the formal $1\sigma$ uncertainties on most of the metal abundances to within a very small range. However, in reality, the true uncertainty on the abundances can be much greater than this due to a number of factors. As demonstrated by Figure \ref{fig:abunplot}, uncertainties in the ionisation balance can lead to order of magnitude differences in the calculated abundance from different ionisation stages. The ionic partition function, which is used to calculate the ionisation balance, is dependent on the atomic data of energy levels, and uncertain data can lead to incorrect calculations of level populations. It is interesting to note that in Figure \ref{fig:abunplot}, the largest difference between ionisation stage abundances was for P and S, where P {\sc iv}/H and P {\sc v}/H differ by $\sim{0.8}$ dex. The best agreement between ion abundances was for C and Si. 

In Figure \ref{fig:abunplot}, we have plotted our measured abundances, comparing them to the solar abundances from \citet{asplund09a}, as well as the atmospheric abundances predicted by \citet{chayer94a,chayer95a,chayer95b} due to radiative levitation. The only obvious agreement that can be observed in Figure \ref{fig:abunplot} is between the radiative levitation prediction and our Fe {\sc v} abundance. The largest discrepancy appears to be for Al, where radiative levitation grossly underestimates the abundance by more than two dex, this may, however, be due to the small number of bound-bound Al transitions included in the calculations. 

\subsection{Opacity considerations}     
The additional atomic data obtained for Fe and Ni has allowed all but a few lines to be identified. However, the number of heavy metal lines also presents a question regarding the efficacy of current methods in accounting for opacity in NLTE model atmosphere calculations. In \citet{chayer95a}, the authors describe calculations performed to predict the abundance of Fe and Ni in the atmosphere of hot white dwarfs due to radiative levitation, and compared the Fe abundances predicted from using different atomic datasets from different generations (see \citealt{chayer95a}, Fig 11), concluding that the numbers of transitions, and the accuracy of their respective oscillator strengths varied the predicted Fe abundances. Such a result implies that using as complete a sample of transitions as possible is important in creating accurate models of radiative levitation. As shown in Table \ref{table:lines}, the number of calculated transitions for Fe {\sc iv}-{\sc vii} alone in 1992 was $\sim{10^7}$, while for 2011 it was $\sim{10^8}$. 
Radiative transfer will also likely be affected by including the additional calculated lines. As described by \citet{barstow98a}, the inclusion of Fe and Ni transitions to a pure H atmosphere results in $T_{\mathrm{eff}}$ decreasing by a few thousand K. In terms of the SED, additional opacity may result in flux attenuation in the EUV, producing better agreement with observation. The additional opacity also potentially offers a solution to the Lyman-Balmer line problem. \cite{barstow03a} constructed two model grids with 0.1 and 10 times a nominal abundance, and measured $T_{\mathrm{eff}}$ and log $g$ for G191-B2B and REJ2214-492 using both the Lyman and Balmer line series. They found that the temperature discrepancy was larger for the lower abundance grid, and smaller for the higher grid. This implies that including additional opacity into future calculations may help to resolve the Lyman-Balmer line problem. To include the additional opacity contributed by the new lines, {\sc tlusty} requires information on the energy levels, transitions, and photoionisation cross sections. {\sc tlusty} is able to calculate bound-bound cross section directly from data provided by the Kurucz data, however, the photoionisation cross sections need to be included separately. Future work will involve calculations of the photoionisation cross sections. Our model atmospheres currently include 2,005,173 Fe {\sc iv-vi} transitions, and 2,751,277 Ni {\sc iv-vi} transitions. Future calculations therefore will include $\sim{30,000,000}$ Fe and $\sim{47,000,000}$ Ni transitions. This work not only has applications to white dwarf stars, but can also be applied to hot O and B stars.

\section{Conclusions}
We have presented the most detailed NUV and FUV spectroscopic survey of G191-B2B to date. Using all available high resolution observations with the E140H and E230H apertures with STIS, we have constructed a co-added spectrum whose S/N exceeds 100 in particular sections of the spectrum, and spans 1160-3145\AA. The detail unveiled by such a high S/N spectrum far exceeds any observation of white dwarfs made thus far. Using the STIS spectrum, and the co-added \textit{FUSE} spectrum from \citet{barstow10a} covering 910-1185\AA, we have made detections of 976 absorption features. By combining the latest releases of the Kurucz and Kentucky databases, we have been able to identify 947 of the detected features, blended or otherwise. We have identified every single absorption feature present in the \textit{FUSE} spectrum of G191-B2B at our given confidence limit. We have found that over 60\% of the identifications made can be attributed to highly ionised Fe and Ni features. While the new Fe and Ni features may be inaccurate in terms of their wavelengths and oscillator strengths, the high S/N STIS spectrum presents the opportunity to perform high quality measurements of atomic transition parameters in order to improve atomic databases.

Our survey confirmed the presence of previously observed circumstellar features and also potentially revealed a new, previously unconsidered Si {\sc iii} circumstellar line.
 
Further, exploiting the high S/N data from the STIS and \textit{FUSE} data, we have made highly accurate measurements of the abundances of metals present in G191-B2B's photosphere. Our abundance analysis has revealed areas for improvement with regards to ionisation balance calculations. S and P appear to have the largest discrepancies.  

We compared the calculated abundances from G191-B2B's atmosphere to the solar photosphere, as well as predicted abundances due to radiative levitation. We highlighted the need for updated radiative levitation calculations using newly available atomic data in order to reconcile theory with observation.

We discussed the potential consequences of including new Fe and Ni transitions into NLTE model atmosphere calculations, as well as radiative levitation calculations. While the exact effect on the NLTE calculations cannot be quantified at present, we plan to perform a full, detailed investigation into the inclusion of new transitions in line blanketing calculations.

\section*{Acknowledgments}

We gratefully thank Jean Dupuis for providing a very thorough referee report, helping to improve the content and quality of this paper.

SPP, MAB, and NJD gratefully acknowledge the support of the Science and Technology Facilities Council (STFC). JBH acknowledges the support of a visiting professorship at the University of Leicester, and the Space Telescope Science Institute Archive Grant AR9202. We gratefully acknowledge the work of Meena Sahu and Wayne Landsman for their work on the original STIS dataset.

We also gratefully acknowledge the help of Peter Van Hoof for assistance in creating the line list used in this paper. This research used the ALICE High Performance Computing Facility at the University of Leicester. All of the spectroscopic data presented in this paper were obtained from the Mikulski Archive for Space Telescopes (MAST). STScI is operated by the Association of Universities for Research in Astronomy, Inc., under NASA contract NAS5-26555. Support for MAST for non-HST data is provided by the NASA Office of Space Science via grant NNX09AF08G and by other grants and contracts

\label{lastpage}
\bibliographystyle{mn2e}

\appendix
\section{Parameterisation of absorption features}
The Gaussian and Lorentzian profiles fitted were parameterised respectively as:
\begin{equation}F_{\lambda}=C_{1}\exp\left[-\frac{(\lambda-C_{2})^2}{2C_{3}^2}\right]\end{equation}
\begin{equation}F_{\lambda}=\frac{C_{1}C_{3}^2}{(\lambda-C_{2})^2+C_{3}^2}\end{equation}
Where $C_1$, $C_2$ and $C_3$ are the height, centroid wavelength and line width respectively. The double Gaussian profile took the form:
\begin{equation}F_{\lambda}=C_{1}\exp\left[-\frac{(\lambda-C_{2})^2}{2C_{3}^2}\right]+C_{4}\exp\left[-\frac{(\lambda-C_{5})^2}{2C_{6}^2}\right]\end{equation}
The $C_{i}$ are the same as defined for a single Gaussian. The Gaussian profile used in {\sc xspec} was parameterised as:
\begin{equation}F(E)=\exp\left[-\frac{\tau_l}{\sigma_{l}\sqrt{2\pi}}\exp\left[-\frac{(E-E_{l})^2}{2\sigma_{l}^{2}}\right]\right]\end{equation}
Where $E_l$ is the line energy in keV (line centroid), $\sigma$ is the line width (keV), and $\tau_{l}$ is the strength. This profile was multiplied by the model flux.

\section{Line identifications}
Included here is a small excerpt of the table of identifications made in our survey. The table in its entirety can be found with the online version of the journal. Wavelengths and their associated errors are given in \AA, equivalent width and error in m\AA, and velocities and errors in km s$^{-1}$. $\lambda_{\mathrm{obs}}$ and $\delta\lambda_{\mathrm{obs}}$, are the observed wavelength centroid and it’s error, and $\lambda_{\mathrm{lab}}$ and $\delta\lambda_{\mathrm{lab}}$ are the reference wavelength and, where available, the error on said wavelength. The velocity $v$ comes with two errors $\delta{v}$ and $\delta{v_{\mathrm{tot}}}$. The former is the error on the velocity assuming no error on the lab wavelength, and the latter takes all errors into account. The List column states where the transition data came from, with KENTUCKY=Kentucky database, KURUCZ=Kurucz database, NIST=NIST website, and RESONANT=V94. The Origin column states where the transition originated. Where PHOT=Photosphere, ISM1=Local Interstellar cloud, and ISM2=Hyades cloud. As was stated previously, many lines had several possible identifications, meaning that a line will be the result of several blends of absorption features. Therefore, where there was more than one identification, the characteristics of the measured line is given, followed only by the additional identifications and their velocities.

\begin{table*}
\centering
\caption{List of detected Photospheric features from \textit{FUSE}. Wavelengths ($\lambda$'s) are given in \AA, wavelength errors in m\AA, equivalent widths in m\AA, and velocities in km s$^{-1}$. The full version of this table has been published online.}
\begin{tabular}[H]{@{}llllllllllll}
\hline
$\lambda_{\mathrm{obs}}$ & $\delta\lambda_{\mathrm{obs}}$ & $W_{\lambda}$ & $\delta{W}_{\lambda}$ & Ion & $\lambda_{\mathrm{lab}}$ & $\delta\lambda_{\mathrm{lab}}$ & $v$ & $\delta{v}$ & $\delta{v_{\mathrm{tot}}}$ & List & Origin \\
\hline
916.503 & 0.533 & 168.903 & 2.369 & H {\sc i}& 916.429 & 0.004 & 24.18 & 0.17 & 0.17 & KENTUCKY & ISM1 \\
917.254 & 0.600 & 170.652 & 2.643 & H {\sc i}& 917.181 & 0.005 & 24.02 & 0.20 & 0.20 & KENTUCKY & ISM1 \\
918.196 & 0.574 & 196.800 & 2.500 & H {\sc i}& 918.129 & 0.007 & 21.78 & 0.19 & 0.19 & KENTUCKY & ISM1 \\
918.964 & 7.564 & 10.266 & 2.198 & N {\sc iv}& 918.893 & 7.500 & 23.16 & 2.47 & 3.48 & KENTUCKY & PHOT \\
919.428 & 0.360 & 178.807 & 1.626 & H {\sc i}& 919.351 & 0.009 & 25.00 & 0.12 & 0.12 & KENTUCKY & ISM1 \\
921.032 & 0.347 & 189.055 & 1.609 & H {\sc i}& 920.963 & 0.012 & 22.45 & 0.11 & 0.11 & KENTUCKY & ISM1 \\
922.090 & 2.485 & 10.595 & 1.404 & N {\sc iv}& 921.994 & 7.600 & 31.22 & 0.81 & 2.60 & KENTUCKY & PHOT \\
922.607 & 2.409 & 10.460 & 1.358 & N {\sc iv}& 922.519 & 7.600 & 28.60 & 0.78 & 2.59 & KENTUCKY & PHOT \\
923.220 & 0.352 & 209.245 & 1.687 & H {\sc i}& 923.150 & 0.016 & 22.64 & 0.11 & 0.11 & KENTUCKY & ISM1 \\
923.757 & 3.069 & 9.727 & 1.454 & N {\sc iv}& 923.676 & 7.600 & 26.29 & 1.00 & 2.66 & KENTUCKY & PHOT \\
\hline
\end{tabular}
\label{table:photfuseident}
\end{table*}

\end{document}